\documentstyle[11pt,amssymb,epsf,fleqn]{article}
\begin{document}
\title{Charged Higgs boson in the next-to-minimal supersymmetric standard model with explicit CP violation}
\author{S.W. Ham$^{(1)}$, J. Kim$^{(1)}$, S.K. Oh$^{(2)}$, and D. Son$^{(1)}$
\\
\\
{\it $^{(1)}$ Center for High Energy Physics, Kyungpook National University} 
\\
{\it Taegu 702-701, Korea} 
\\
{\it $^{(2)}$ Department of Physics, Konkuk University, Seoul 143-701, 
Korea}
\\
\\
}
\date{}
\maketitle
\begin{abstract}
The phenomenology of the explicit CP violation in the Higgs sector of the next-to-minimal supersymmetric standard model (NMSSM) is investigated, with emphasis on the charged Higgs boson.
The radiative corrections due to both quarks and scalar-quarks of the third generation are taken into account, and the negative result of the search for the Higgs bosons at CERN LEP2, with the discovery limit of 0.1 pb, is imposed as a constraint.
It is found that there are parameter regions of the NMSSM where the lightest neutral Higgs boson may even be massless, without being detected at LEP2.
This implies that the LEP2 data do not contradict the existence of a massless neutral Higgs boson in the NMSSM.
For the charged Higgs boson, the radiative corrections to its mass may be negative in some parameter regions of the NMSSM. 
The phenomenological lower bound on the radiatively corrected mass of the charged Higgs boson is increased as the CP violation becomes maximal, i.e., as the CP violating phase becomes $\pi/2$. 
At the maximal CP violation, its lower bound is about 110 GeV for 5 $\leqslant \tan \beta \leqslant$ 40.
The vacuum expectation value (VEV) of the neutral Higgs singlet is shown to be 
no smaller than 16 GeV for any parameter values of the NMSSM with explicit CP violation. 
This value of the lower limit is found to increase up to about 45 GeV as the ratio ($\tan \beta$) of the VEVs of the two Higgs doublets decreases to smaller values ($\sim$ 2).
The discovery limit of the Higgs boson search at LEP2 is found to cover about a half of the kinematically allowed part of the whole parameter space of the NMSSM, and the portion is roughly stable against the CP violating phase.
\end{abstract}
\vfil

%***********************************************************************
\section{INTRODUCTION}
%***********************************************************************

The violation of CP symmetry has been with us for more than three decades since it was first observed in the neutral kaon system [1].
Within the context of the standard model (SM) [2] the CP violation is accommodated by virtue of a complex phase that exists in the Cabibbo-Kobayashi-Maskawa (CKM) matrix for the charged weak current [3]. 
The origin of the complex phase has been remained in the standard
model unexplained.
A number of speculations have been proposed for possible explanations.
One of the possibilities is that CP symmetry is violated in the Higgs sector, if there are at least two Higgs doublets [4].
It is known that supersymmetric standard models require at least two Higgs doublets in their Higgs sectors.
The two Higgs doublets give separately masses to the up-type quarks and the down-type quarks [5].
If the supersymmetry (SUSY) is exact in the model, where the SM particles and their superparticles have equal masses, CP violation would be absent in its Higgs sector, since certain restrictions from the SUSY are imposed on the Higgs sector.
The supersymmetric model cannot however have an exact SUSY since the nature is not supersymmetric at electroweak level.
If the SUSY is broken somehow in the model for phenomenological analyses, for example by introduction of the soft SUSY breaking terms, CP violation may arise in the Higgs sector of the supersymmetric standard model.

In the minimal supersymmetric standard model (MSSM), many authors have investigated the possibilities of CP violation either explicitly or spontaneously. 
It is found that the tree-level Higgs potential of the MSSM is unable to break the CP symmetry in either way. 
Further investigations at 1-loop level of the MSSM show that the scenario of spontaneous CP violation is experimentally ruled out by the Higgs boson serch at the CERN $e^+ e^-$ LEP, because it leads to a very light neutral Higgs boson [6].
However, at the 1-loop level, the scenario of explicit CP violation in the MSSM is found to be viable through the radiatively corrected Higgs potential [7].
Subsequently, a number of investigations have been devoted to the effects of explicit CP violation in the MSSM at 1-loop level to the charged Higgs boson [8] as well as to the neutral ones [9] by considering radiative corrections. 
It has been expected that at higher level the effect of explicit CP violation in the MSSM may lead to significant modifications to the tree level Higgs couplings to fermions and to gauge bosons.

Meanwhile, the CP symmetry may also be broken if the Higgs sector is extended by introducing at least one Higgs singlet. 
Then, the next-to-minimal supersymmetric standard model (NMSSM) is the right candidate for CP violation, which is basically different from the MSSM in that it has one extra Higgs singlet.
For the spontaneous CP violation, it is well known that the tree-level Higgs potential of the NMSSM cannot provide the ground [10]. 
The situation is not improved at the 1-loop level by considering radiative corrections, if the top scalar quarks are degenerated in their masses [11].
In the NMSSM Higgs sector, spontaneous CP violation may only be realized by virtue of radiative corrections when the mass splitting effects between the top scalar-quarks are taken into account [12].

At the tree-level, unlike the MSSM, the scenario of explicit CP violation is acceptable by the NMSSM. 
By assuming the degeneracy of the top scalar-quark masses in the Higgs sector of the NMSSM, it is found that large explicit CP violation may be realized as the vacuum expectation value (VEV) of the neutral Higgs singlet approaches to the electroweak scale [13].  

In this paper we investigate the effects of the explicit CP violation on the charged Higgs boson in the NMSSM.
We calculate radiative corrections to the masses of both the charged Higgs boson and the neutral ones in the NMSSM due to the quarks and scalar-quarks of the third generation. 
The 1-loop effective potential which includes the effects of the mass splittings in the scalar-quarks is employed.
Assuming the nondegeneracy of the scalar-quark masses of the third generation in the 1-loop effective potential, a CP violating phase occurs from soft SUSY breaking terms, which penetrates into the radiatively corrected masses of the charged Higgs boson as well as of the neutral ones. 
Phenomenological implications and constraints by those CP violating Higgs masses are investigated over a wide region of the parameter space of the NMSSM, using the production processes of the neutral Higgs boson in $e^+ e^-$ collisions.
In particular, the results at LEP2 with $\sqrt{s}$=200 GeV are used.

Our paper is organized as follows. 
In the next section, we consider radiative corrections to the Higgs boson mass by using the 1-loop effective potential in explicit CP violation scenario.
In the third section, we obtain phenomenological constraints on the charged Higgs boson mass, the VEV of the neutral Higgs singlet, by considering the neutral Higgs production at LEP2.
In the last section, conclusions are discussed.

%***********************************************************************
\section{HIGGS SECTOR}
%***********************************************************************

The Higgs sector of the NMSSM contains two Higgs doublet superfields $H_1^T$ = ($H_1^0$, $H_1^-$), $H_2^T$ = ($H_2^+$, $H_2^0$), and a neutral Higgs singlet superfield $N$.
Ignoring all quark and lepton Yukawa couplings except for those of the third generation quarks, the superpotential of the NMSSM including only terms with dimensionless couplings is written as [14]
\[
W = h_t Q H_2 t_R^c + h_b Q H_1 b_R^c + \lambda N H_1 H_2 - {k \over 3} N^3 \ ,
\]
where $h_t$ and $h_b$ are Yukawa coupling constants of the third generation quarks.
The parameters $\lambda$ and $k$ are dimensionless coupling constants.
We define that $H_1 H_2$ = $H_1^0 H_2^2 - H_1^- H_2^+$.
The SU(2) doublet superfield $Q^T$ = ($t_L$, $b_L$) contains the left-handed quarks for the third generation.
The SU(2) singlet superfields $t_R^c$ and $b_R^c$ are the charge conjugates of the right-handed quarks for the third generation.
A global U(1) Peccei-Quinn symmetry would be explicitly broken by the presence of the cubic term of the Higgs singlet in the superpotential, which leads us to have a massless pseudoscalar Higgs boson at the tree level.

At the tree level the Higgs potential of the model consist of $F$ terms, $D$ terms, and soft terms as
\[
    V^0 = V_{\rm F} + V_{\rm D} + V_{\rm soft}  \ , 
\]
where
\begin{eqnarray}
    V_{\rm F} &=& |\lambda|^2[(|H_1|^2+|H_2|^2)|N|^2+|H_1 H_2|^2]
            + |k|^2|N|^4 -(\lambda k^*H_1H_2N^{*2}+ {\rm H.c.}) \ , \cr
    V_{\rm D} &=& {1 \over 8} (g_1^2 + g_2^2) (|H_2|^2-|H_1|^2)^2 \ , \\
    V_{\rm soft} &=& m_{H_1}^2|H_1|^2+m_{H_2}^2|H_2|^2+m_N^2|N|^2
     - (\lambda A_\lambda H_1 H_2 N + {1\over 3} k A_k N^3+ {\rm H.c.})
     \ , \nonumber
\end{eqnarray}
with $g_1$ and $g_2$ being the U(1) and SU(2) gauge coupling constants, respectively.
In the soft-terms, $m_{H_1}^2$, $m_{H_2}^2$, and $m_N^2$ are the soft SUSY breaking masses.
$A_{\lambda}$ and $A_k$ are the trilinear soft SUSY breaking parameters with mass dimension.
The role of the soft-terms in the Higgs potential is to break SUSY.
The $F$ terms come from the part of chiral auxiliary superfield while the $D$ terms come from the part of gauge auxiliary superfield.

In order to consider the explicit CP violation in the Higgs sector, $\lambda$, $A_{\lambda}$, $k$, and $A_k$ are assumed to be complex numbers.
Among them, $\lambda A_{\lambda}$ and $k A_k$ can be adjusted to be real and positive by redefining the overall phases of the superfields $H_1 H_2$ and $N$.
Then, we are left essentially with one phase: It may be chosen to be the one in $\lambda k^*$ = $\lambda k e^{i \phi}$, which allows for explicit CP violation in the tree level Higgs potential of the NMSSM.
We assume that the vacuum expectation values (VEVs) of three neutral Higgs fields $H_1^0$, $H_2^0$, and $N$ are nonvanishing values, both real and positive, so that there is no spontaneous CP violation in the Higgs sector.

The conditions that the Higgs potential becomes minimum at $v_1 = \langle H_1^0 \rangle$, $v_2 = \langle H_2^0 \rangle$, and $x = \langle N \rangle$ yield three constraints, which can eliminate the soft SUSY breaking masses $m_{H_1}$, $m_{H_2}$, and $m_N$ from the potential $V^0$.
Note that although the phase $\phi$ is present in the potential, it disappears from the vacuum after applying the three minimum conditions for the VEVs. 
Thus, one does not impose a minimum condition on the Higgs potential with respect to the CP violating phase $\phi$.
At the tree level, in consequence, the Higgs potential contains $\phi$, $\tan \beta=v_2/v_1$, $\lambda$, $A_{\lambda}$, $k$, $A_k$, and $x$ as free parameters.

The Higgs sector of the NMSSM has ten real degrees of freedom coming from two Higgs doublets and one Higgs singlet.
After the electroweak symmetry breakdown takes place, three of the ten components correspond to a neutral Goldstone boson and a pair of charged Goldstone bosons, the other seven correspond to five neutral Higgs bosons and a pair of charged Higgs bosons.
Transforming to a unitary gauge, one may express two Higgs doublets and one Higgs singlet as
\begin{equation}
\begin{array}{ccc}
H_1 & = & \left ( \begin{array}{c}
          v_1 + S_1 + i \sin \beta A   \cr
          \sin \beta C^{+ *}
  \end{array} \right )  \ ,  \cr
H_2 & = & \left ( \begin{array}{c}
          \cos \beta C^+           \\
          v_2 + S_2 + i \cos \beta A
  \end{array} \right ) \ ,   \\
N & = & ( \begin{array}{c}
          x + X + i Y
        \end{array} )  \ , \nonumber
\end{array}
\end{equation}
where $S_1$, $S_2$, $A$, $X$, and $Y$ are neutral fields and $C^+$ is the charged field.

If CP symmetry be conserved in the Higgs potential, $S_1$, $S_2$, and $X$ would become the CP even scalar Higgs fields whereas $A$ and $Y$ would become the CP odd pseudoscalar Higgs fields.
The three orthogonal combinations among $A$ and $C^{\pm}$ yield one neutral Goldstone boson and a pair of charged Goldstone bosons; they will eventually be absorbed into the longitudinal component of $Z$ and $W$ gauge bosons, respectively.
We take the constraints $m_W^2 = g_2^2 v^2/2$ and $m_Z^2 = (g_1^2 + g_2^2) v^2/2$ for the electroweak gauge bosons with $v = \sqrt{v_1^2 + v_2^2}$ = 175 GeV.

It has already been observed that the above tree level Higgs potential is significantly affected by radiative corrections due to the quarks and scalar-quarks of the third generation.
Thus, we also take into account, in the presence of explicit CP violation, the radiative corrections. 
We employ here the method of the effective potential.
The 1-loop effective potential including radiative corrections due to the quark and scalar-quark of the third generation is given by [15]
\[
V^1 = \frac{3}{32\pi^2} \left \{ {\cal M}_{\tilde{q_i}}^4
   \left (\log {{\cal M}_{\tilde{q_i}}^2 \over \Lambda^2} - {3\over 2} \right )
   - 2 {\cal M}_q^4 \left (\log {{\cal M}_q^2 \over \Lambda^2}
   - {3\over 2} \right ) \right \}  \ ,
\]
where ${\cal M}_{{\tilde q}_i}^2$ ($i$=1 to 4) are the field dependent scalar-quark masses for the third generation, ${\cal M}_q^2$ ($q$ = $b, t$) are the field dependent quark masses for the third generation, and $\Lambda$ is the renormalization scale in the modified minimal subtraction $\overline {\rm MS}$ scheme.
Note that the scalar-quark term is positive in the 1-loop effective potential whereas the quark term is negative.
Because of the relative opposite sign between the scalar quarks and the quarks, if masses of the quarks and their corresponding scalar quarks are equal, then there would be no net contribution from the radiative corrections. 
However, this should not be the case since no light scalar particle with mass comparable to light quarks has been observed: The quarks and scalar-quarks should not have equal masses.

The top quark and bottom quark masses are given as $m_t^2 = (h_t v_2)^2$ from $h_t^2 (|H_2^0|^2 + |H_2^+|^2)$ and as $m_b^2 = (h_b v_1)^2$ from $h_b^2 (|H_1^0|^2 + |H_1^-|^2)$ after the electroweak symmetry breakdown, respectively.
By taking the basis (${\tilde t}_L$, ${\tilde t}_R^c$, ${\tilde b}_L$, ${\tilde b}_R^c$) for the scalar-quark mass matrix of the third generation in Lagrangian density, the elements of the 4 $\times$ 4 Hermitian matrix, ${\cal M}_{\tilde q}$, are explicitly given as
\begin{eqnarray}
 {\cal M}_{11} & = & m_Q^2 + h_t^2 |H_2^0|^2 + h_b^2 |H_1^-|^2  \ , \cr
 {\cal M}_{22} & = & m_T^2 + h_t^2 |H_2^0|^2 + h_t^2 |H_2^+|^2  \ , \cr
 {\cal M}_{33} & = & m_Q^2 + h_b^2 |H_1^0|^2 + h_t^2 |H_2^+|^2  \ , \cr
 {\cal M}_{44} & = & m_B^2 + h_b^2 |H_1^0|^2 + h_b^2 |H_1^-|^2  \ , \cr
 {\cal M}_{12} & = & h_t (\lambda N H_1^0 + A_t H_2^{0 *})  \ , \cr
 {\cal M}_{13} & = & - h_t^2 H_2^{0 *} H_2^+ - h_b^2 H_1^0 H_1^{- *} \ , \\
 {\cal M}_{14} & = & - h_b \lambda N H_2^+ + h_b A_b H_1^{- *} \ , \cr
 {\cal M}_{23} & = & h_t \lambda N^* H_1^{- *} - h_t A_t H_2^+ \ , \cr
 {\cal M}_{24} & = & h_t h_b H_2^0 H_1^{- *} + h_t h_b H_1^{0 *} H_2^+ \ , \cr
 {\cal M}_{34} & = & - h_b \lambda N H_2^0 - h_b A_b H_1^{0 *} \ , \nonumber
\end{eqnarray}
where $m_Q^2$, $m_T^2$, and $m_B^2$ are soft SUSY breaking masses and $A_q$ ($q = t, b$) are the trilinear soft SUSY breaking parameters.
In the above expressions, the contribution from the $D$ terms are not included.

In order to accommodate the possibility of explicit CP violation at the 1-loop level, some of the coupling coefficients in the 1-loop effective potential are assumed to be complex.
Note first that the complex phases of both $h_t$ and $h_b$ can later be absorbed away; they are thus assumed to be real. 
Next, we assume that $m_T = m_B$ and $A_t = A_b$, on the ground that the leading radiative corrections come from the contribution of the top quark and top scalar-quark, although those of the bottom quark and bottom scalar quark for large $\tan \beta$ is not negligible.
Accordingly, we may set the phase of $A_t$ to be equal to that of $A_b$.
Consequently, at the 1-loop level, there remains only one relative phase between $A_t = A_b$ and $\lambda$. 
We name it as $\phi_1 = \phi_{\lambda} - \phi_{A_q}$, and assume for simplicity that it is equal to the CP violating phase at tree level: $\phi_1 = \phi$.

The scalar-quark mass matrix for the third generation may conveniently be rewritten into four 2 $\times$ 2 block submatrices. 
Among them, the two off-diagonal block submatrices become zero matrices after the electroweak symmetry breakdown, since they do not contain the neutral Higgs fields.
The two diagonal block submatrices may have non-zero mass eigenvalues, by substituting the VEVs for the neutral Higgs fields. 
Actually, by diagonalizing each of the block submatrices, one can obtain the top scalar-quark masses $m_{{\tilde q}_1}^2$ and $m_{{\tilde q}_2}^2$ ($m_{{\tilde q}_1}^2 < m_{{\tilde q}_2}^2$) as
\begin{equation}
  m_t^2 + {1 \over 2}(m_Q^2 + m_T^2) \mp
\sqrt{{1 \over 4} (m_Q^2 - m_T^2)^2 + m_t^2 (A_t^2 + \lambda^2 x^2 \cot^2 \beta
+ 2 A_t \lambda x \cot \beta \cos \phi)} \ , 
\end{equation}
and the bottom scalar-quark masses $m_{{\tilde q}_3}^2$ and $m_{{\tilde q}_4}^2$($m_{{\tilde q}_3}^2 < m_{{\tilde q}_4}^2$) as 
\begin{equation}
 m_b^2 + {1 \over 2} (m_Q^2 + m_T^2) \mp
\sqrt{{1 \over 4} (m_Q^2 - m_T^2)^2 + m_b^2 (A_t^2 + \lambda^2 x^2 \tan^2 \beta
+ 2 A_t \lambda x \tan \beta \cos \phi)} \ . 
\end{equation}
It is clearly the phase $\phi$ in the scalar-quark masses that is the possible source of the explicit CP violation.
Notice that at the 1-loop level the phase $\phi$ of the explicit CP violation would be absent as long as the scalar-quarks of the third generation are degenerate in their masses.
We assume that the left-handed and the right-handed scalar quark masses are not degenerate. 

The full Higgs potential of the NMSSM up to the 1-loop level may be decomposed into the neutral and the charged parts, according to which Higgs fields are involved.
Then, the neutral part of the Higgs potential would yield the 5 $\times$ 5 mass matrix for the five neutral Higgs bosons, if it is differentiated twice with respect to the corresponding neutral Higgs fields. 
The charged part would also yield charged Higgs mass.

Let us consider first the 5 $\times$ 5 mass matrix for the five neutral Higgs bosons. 
The elements of the matrix can be expressed, at the basis of ($S_1$, $S_2$, $A$, $X$, $Y$), as
\begin{equation}
M_{ij} = M_{ij}^0 + \delta M_{ij}^t + \delta M_{ij}^b 
\ (i, \ j=1 \ {\rm to} \ 5) \ ,
\end{equation}
where $M_{ij}^0$ come from the tree level Higgs potential, $\delta M_{ij}^t$ from the top quark and top scalar-quark contributions, and $\delta M_{ij}^b$ from the bottom quark and bottom scalar-quark contributions.
We calculate them in an explicit manner to obtain the following results.
First, the tree level Higgs potential gives 
%*********************  tree level ******************************
\begin{eqnarray}
M_{11}^0 & = & (m_Z \cos \beta)^2
+ \lambda x (A_{\lambda} + kx \cos \phi) \tan \beta  \ , \cr
M_{22}^0 & = & (m_Z \sin \beta)^2
+ \lambda x (A_{\lambda} + kx \cos \phi) \cot \beta  \ , \cr
M_{33}^0 & = & {2 \lambda x (A_{\lambda}
+ kx \cos \phi) \over \sin 2 \beta}  \ , \cr
M_{44}^0 & = & (2 k x)^2 - k x A_k
+ {\lambda \over 2x} v^2 A_{\lambda} \sin 2 \beta \ , \cr
M_{55}^0 & = & {\lambda v^2 \over 2x} A_{\lambda} \sin 2 \beta
+ 3 k x A_k + 2 \lambda k v^2 \sin 2 \beta \cos \phi  \ , \cr
M_{12}^0 & = & (\lambda^2 v^2 - {1 \over 2} m_Z^2) \sin 2 \beta
- \lambda x (A_{\lambda} + k x \cos \phi) \ , \cr
M_{13}^0 & = & 0 \ , \cr
M_{14}^0 & = & 2 \lambda^2 x v \cos \beta - \lambda v \sin \beta (A_{\lambda}
+ 2 k x \cos \phi) \ , \cr
M_{15}^0 & = &\mbox{} - 3 \lambda k v x \sin \beta \sin \phi \ , \cr
M_{23}^0 & = & 0 \ , \cr
M_{24}^0 & = & 2 \lambda^2 x v \sin \beta
- \lambda v \cos \beta (A_{\lambda} + 2 k x \cos \phi) \ , \cr
M_{25}^0 & = &\mbox{} - 3 \lambda k v x \cos \beta \sin \phi \ , \cr
M_{34}^0 & = &\mbox{} \lambda k v x \sin \phi \ , \cr
M_{35}^0 & = &\mbox{} \lambda v (A_{\lambda} - 2 k x \cos \phi) \ , \cr
M_{45}^0 & = & 2 \mbox{} \lambda k v^2 \sin 2 \beta \sin \phi \ , \nonumber
\end{eqnarray}
and the contributions of the top quark and top scalar-quark give
%*********************  top and stop ******************************
\begin{eqnarray}
 \delta M_{11}^t & = &
{3 m_t^4 \lambda^2 x^2 \Delta_{{\tilde t}_1}^2
\over 8 \pi^2  v^2 \sin^2 \beta}
{g(m_{\tilde{t}_1}^2, \ m_{\tilde{t}_2}^2) \over
(m_{\tilde{t}_2}^2 - m_{\tilde{t}_1}^2)^2}
+ {3 m_t^2 \lambda x A_t \cos \phi
\over 8 \pi^2 v^2 \sin 2 \beta}
f(m_{\tilde{t}_1}^2, \ m_{\tilde{t}_2}^2)   \ , \cr
 & & \cr
%%%%%%%%%
 \delta M_{22}^t & = &
{3 m_t^4 A_t^2 \Delta_{{\tilde t}_2}^2
\over 8 \pi^2  v^2 \sin^2 \beta}
{g(m_{\tilde{t}_1}^2, \ m_{\tilde{t}_2}^2)
\over (m_{\tilde{t}_2}^2 - m_{\tilde{t}_1}^2)^2}
+ {3 m_t^2 \lambda x A_t \cos \phi
\over 16 \pi^2 v^2  \sin^2 \beta \tan \beta}
f(m_{\tilde{t}_1}^2, \ m_{\tilde{t}_2}^2) \cr
 & &\mbox{}
+ {3 m_t^4 A_t \Delta_{{\tilde t}_2}
\over 4 \pi^2 v^2 \sin^2 \beta}
{\log (m_{\tilde{t}_2}^2 / m_{\tilde{t}_1}^2)
 \over (m_{\tilde{t}_2}^2 - m_{\tilde{t}_1}^2)}
+ {3 m_t^4 \over 8 \pi^2 v^2 \sin^2 \beta}
\log \left ( {m_{\tilde{t}_1}^2  m_{\tilde{t}_2}^2 \over m_t^4} \right ) \ , \cr
 & & \cr
%%%%%%%%%
 \delta M_{33}^t & = & {3 m_t^4 \lambda^2 x^2 A_t^2 \sin^2 \phi
\over 8 \pi^2 v^2 \sin^4 \beta}
{g(m_{\tilde{t}_1}^2, \ m_{\tilde{t}_2}^2) \over (m_{\tilde{t}_2}^2
- m_{\tilde{t}_1}^2 )^2}
+ {3 m_t^2 \lambda x A_t \cos \phi \over 16 \pi^2 v^2 \sin^3 \beta \cos \beta}
f(m_{\tilde{t}_1}^2, \ m_{\tilde{t}_2}^2)  \ , \cr
 & & \cr
%%%%%%%%%
 \delta M_{44}^t & = &
{3 m_t^4 \lambda^2 \Delta_{{\tilde t}_1}^2 \over 8 \pi^2 \tan^2 \beta}
{g(m_{\tilde{t}_1}^2, \ m_{\tilde{t}_2}^2)
\over (m_{\tilde{t}_2}^2 - m_{\tilde{t}_1}^2 )^2}
+ {3 m_t^2 \lambda A_t \cos \phi \over 16 \pi^2 x \tan \beta}
f(m_{\tilde{t}_1}^2, \ m_{\tilde{t}_2}^2)  \ , \cr
 & & \cr
%%%%%%%%%
 \delta M_{55}^t & = & {3 m_t^4 \lambda^2 A_t^2 \sin^2 \phi
\over 8 \pi^2 \tan^2 \beta}
{g(m_{\tilde{t}_1}^2, \ m_{\tilde{t}_2}^2) \over (m_{\tilde{t}_2}^2
- m_{\tilde{t}_1}^2 )^2}
+ {3 m_t^2 \lambda A_t \cos \phi
\over 16 \pi^2 x \tan \beta}
f(m_{\tilde{t}_1}^2, \ m_{\tilde{t}_2}^2)  \ ,  \cr
 & & \cr
%%%%%%%%%
 \delta M_{12}^t & = &
{3 m_t^4 \lambda x A_t \Delta_{{\tilde t}_1} \Delta_{{\tilde t}_2}
\over 8 \pi^2 v^2 \sin^2 \beta}
{g(m_{\tilde{t}_1}^2, \ m_{\tilde{t}_2}^2)
\over (m_{\tilde{t}_2}^2 - m_{\tilde{t}_1}^2)^2}
- {3 m_t^2 \lambda x A_t \cos \phi \over 16 \pi^2 v^2 \sin^2 \beta}
f(m_{\tilde{t}_1}^2, \ m_{\tilde{t}_2}^2)  \cr
 & &\mbox{}
+ {3 m_t^4 \lambda x \Delta_{{\tilde t}_1} \over 8 \pi^2 v^2 \sin^2 \beta}
{\displaystyle \log (m_{\tilde{t}_2}^2 / m_{\tilde{t}_1}^2)
 \over (m_{\tilde{t}_2}^2 - m_{\tilde{t}_1}^2)} \cr
 & & \cr
%%%%%%%%%
 \delta M_{13}^t & = & \mbox{}
- {3 m_t^4 \lambda^2 x^2 A_t \sin \phi \Delta_{{\tilde t}_1}
\over 8 \pi^2 v^2 \sin^3 \beta}
{g(m_{\tilde{t}_1}^2, \ m_{\tilde{t}_2}^2) \over
(m_{\tilde{t}_2}^2 - m_{\tilde{t}_1}^2)^2 }
+ {3 m_t^2 \lambda x A_t \sin \phi \over 16 \pi^2 v^2 \sin \beta \tan \beta}
f(m_{\tilde{t}_1}^2, \ m_{\tilde{t}_2}^2) \ , \cr
 & & \cr
%%%%%%%%%
 \delta M_{14}^t & = &
{3 m_t^4 \lambda^2 x \Delta_{{\tilde t}_1}^2
\over 8 \pi^2 v \sin \beta \tan \beta}
{g(m_{\tilde{t}_1}^2, \ m_{\tilde{t}_2}^2)
\over (m_{\tilde{t}_2}^2 - m_{\tilde{t}_1}^2)^2 }
- {3 m_t^2 \lambda (\Delta_{{\tilde t}_1} + \lambda x \cot \beta)
\over 16 \pi^2 v \sin \beta}
f(m_{\tilde{t}_1}^2, \ m_{\tilde{t}_2}^2)  , \cr
 & & \cr
%%%%%%%%
 \delta M_{15}^t & = &\mbox{}
- {3 m_t^4 \lambda^2 x A_t \sin \phi \Delta_{{\tilde t}_1}
\over 8 \pi^2 v \sin \beta \tan \beta}
{g(m_{\tilde{t}_1}^2, \ m_{\tilde{t}_2}^2) \over
(m_{\tilde{t}_2}^2 - m_{\tilde{t}_1}^2)^2 }
+ {3 m_t^2 \lambda A_t \sin \phi \over 16 \pi^2 v \sin \beta \tan \beta}
f(m_{\tilde{t}_1}^2, \ m_{\tilde{t}_2}^2) \ , \cr
 & & \cr
%%%%%%%%%
 \delta M_{23}^t & = & \mbox{}
- {3 m_t^4 \lambda x A_t^2 \sin \phi \Delta_{{\tilde t}_2}
\over 8 \pi^2 v^2 \sin^3 \beta}
{g(m_{\tilde{t}_1}^2, \ m_{\tilde{t}_2}^2) \over
(m_{\tilde{t}_2}^2 - m_{\tilde{t}_1}^2)^2 }
+ {3 m_t^2 \lambda x A_t \sin \phi \over 16 \pi^2 v^2 \sin \beta}
f(m_{\tilde{t}_1}^2, \ m_{\tilde{t}_2}^2) \cr
 & &\mbox{}
- {3 m_t^4 \lambda x A_t \sin \phi \over 8 \pi^2 v^2 \sin^3 \beta}
{\log (m_{\tilde{t}_2}^2 / m_{\tilde{t}_1}^2) \over
(m_{\tilde{t}_2}^2 - m_{\tilde{t}_1}^2)}  \ , \cr
 & & \cr
%%%%%%%%%
 \delta M_{24}^t & = &
{3 m_t^4 \lambda A_t \Delta_{{\tilde t}_1} \Delta_{{\tilde t}_2}
\over 8 \pi^2 v \sin \beta \tan \beta}
{g(m_{\tilde{t}_1}^2, \ m_{\tilde{t}_2}^2) \over
(m_{\tilde{t}_2}^2 - m_{\tilde{t}_1}^2)^2}
- {3 m_t^2 \lambda A_t \cos \phi \over 16 \pi^2 v \sin \beta \tan \beta}
f(m_{\tilde{t}_1}^2, \ m_{\tilde{t}_2}^2) \cr
 & &\mbox{} + {3 m_t^4 \lambda \Delta_{{\tilde t}_1}
\over 8 \pi^2 v \sin \beta \tan \beta}
{\log (m_{\tilde{t}_2}^2 / m_{\tilde{t}_1}^2)
\over (m_{\tilde{t}_2}^2 - m_{\tilde{t}_1}^2) }  , \cr
 & & \cr
%%%%%%%%%
 \delta M_{25}^t & = & \mbox{}
- {3 m_t^4 \lambda A_t^2 \sin \phi \Delta_{{\tilde t}_2}
\over 8 \pi^2 v \sin \beta \tan \beta}
{g(m_{\tilde{t}_1}^2, \ m_{\tilde{t}_2}^2) \over
(m_{\tilde{t}_2}^2 - m_{\tilde{t}_1}^2)^2 }
+ {3 m_t^2 \lambda A_t \sin \phi \over 16 \pi^2 v \sin \beta \tan \beta}
f(m_{\tilde{t}_1}^2, \ m_{\tilde{t}_2}^2) \cr
 & &\mbox{}
- {3 m_t^4 \lambda A_t \sin \phi \over 8 \pi^2 v \sin \beta \tan \beta}
{ \log (m_{\tilde{t}_2}^2 / m_{\tilde{t}_1}^2) \over
(m_{\tilde{t}_2}^2 - m_{\tilde{t}_1}^2)}  \ , \cr
 & & \cr
%%%%%%%%%
 \delta M_{34}^t & = & \mbox{}
- {3 m_t^4 \lambda^2 x A_t \sin \phi \Delta_{{\tilde t}_1}
\over 8 \pi^2 v \sin^2 \beta \tan \beta}
{g(m_{\tilde{t}_1}^2, \ m_{\tilde{t}_2}^2)
\over (m_{\tilde{t}_2}^2 - m_{\tilde{t}_1}^2)^2 }
+ {3 m_t^2 \lambda A_t \sin \phi \over 16 \pi^2 v \sin^2 \beta}
f(m_{\tilde{t}_1}^2, \ m_{\tilde{t}_2}^2)  \ , \cr
 & & \cr
%%%%%%%%%
 \delta M_{35}^t & = & \mbox{}
{3 m_t^4 \lambda^2 x A_t^2 \sin^2 \phi
\over 8 \pi^2 v \sin^2 \beta \tan \beta}
{g(m_{\tilde{t}_1}^2, \ m_{\tilde{t}_2}^2)
\over (m_{\tilde{t}_2}^2 - m_{\tilde{t}_1}^2)^2 }
+ {3 m_t^2 \lambda A_t \cos \phi \over 16 \pi^2 v \sin^2 \beta}
f(m_{\tilde{t}_1}^2, \ m_{\tilde{t}_2}^2)  \ , \cr
 & & \cr
%%%%%%%%%
 \delta M_{45}^t & = & \mbox{}
- {3 m_t^4 \lambda^2 A_t \sin \phi \Delta_{{\tilde t}_1}
\over 8 \pi^2 \tan^2 \beta}
{ g(m_{\tilde{t}_1}^2, \ m_{\tilde{t}_2}^2)
\over (m_{\tilde{t}_2}^2 - m_{\tilde{t}_1}^2)^2 } \ ,
\end{eqnarray}
and finally the contributions of the bottom quark and bottom scalar-quark give
%*********************  bottom and sbottom ******************************
\begin{eqnarray}
 \delta M_{11}^b & = &
{3 m_b^4 A_t^2 \Delta_{{\tilde b}_2}^2
\over 8 \pi^2  v^2 \cos^2 \beta}
{g(m_{\tilde{b}_1}^2, \ m_{\tilde{b}_2}^2)
\over (m_{\tilde{b}_2}^2 - m_{\tilde{b}_1}^2)^2}
+ {3 m_b^2 \lambda x A_t \cos \phi
\over 16 \pi^2 v^2  \cos^2 \beta \cot \beta}
f(m_{\tilde{b}_1}^2, \ m_{\tilde{b}_2}^2) \cr
 & &\mbox{}
+ {3 m_b^4 A_t \Delta_{{\tilde b}_2}
\over 4 \pi^2 v^2 \cos^2 \beta}
{\log (m_{\tilde{b}_2}^2 / m_{\tilde{b}_1}^2)
 \over (m_{\tilde{b}_2}^2 - m_{\tilde{b}_1}^2)}
+ {3 m_b^4 \over 8 \pi^2 v^2 \cos^2 \beta}
\log \left ( {m_{\tilde{b}_1}^2  m_{\tilde{b}_2}^2 \over m_b^4} \right ) \ , \cr
 & & \cr
%%%%%%%%%
 \delta M_{22}^b & = &
{3 m_b^4 \lambda^2 x^2 \Delta_{{\tilde b}_1}^2
\over 8 \pi^2  v^2 \cos^2 \beta}
{g(m_{\tilde{b}_1}^2, \ m_{\tilde{b}_2}^2) \over
(m_{\tilde{b}_2}^2 - m_{\tilde{b}_1}^2)^2}
+ {3 m_b^2 \lambda x A_t \cos \phi
\over 8 \pi^2 v^2 \sin 2 \beta}
f(m_{\tilde{b}_1}^2, \ m_{\tilde{b}_2}^2)   \ , \cr
 & & \cr
%%%%%%%%%
 \delta M_{33}^b & = & {3 m_b^4 \lambda^2 x^2 A_t^2 \sin^2 \phi
\over 8 \pi^2 v^2 \cos^4 \beta}
{g(m_{\tilde{b}_1}^2, \ m_{\tilde{b}_2}^2) \over (m_{\tilde{b}_2}^2
- m_{\tilde{b}_1}^2 )^2}
+ {3 m_b^2 \lambda x A_t \cos \phi
\over 16 \pi^2 v^2 \cos^3 \beta \sin \beta}
f(m_{\tilde{b}_1}^2, \ m_{\tilde{b}_2}^2)  \ , \cr
 & & \cr
%%%%%%%%%
 \delta M_{44}^b & = &
{3 m_b^4 \lambda^2 \Delta_{{\tilde b}_1}^2 \over 8 \pi^2 \cot^2 \beta}
{g(m_{\tilde{b}_1}^2, \ m_{\tilde{b}_2}^2)
\over (m_{\tilde{b}_2}^2 - m_{\tilde{b}_1}^2 )^2}
+ {3 m_b^2 \lambda A_t \cos \phi \over 16 \pi^2 x \cot \beta}
f(m_{\tilde{b}_1}^2, \ m_{\tilde{b}_2}^2)  \ , \cr
 & & \cr
%%%%%%%%%
 \delta M_{55}^b & = & {3 m_b^4 \lambda^2 A_t^2 \sin^2 \phi
\over 8 \pi^2 \cot^2 \beta}
{g(m_{\tilde{b}_1}^2, \ m_{\tilde{b}_2}^2) \over (m_{\tilde{b}_2}^2
- m_{\tilde{b}_1}^2 )^2}
+ {3 m_b^2 \lambda A_t \cos \phi
\over 16 \pi^2 x \cot \beta}
f(m_{\tilde{b}_1}^2, \ m_{\tilde{b}_2}^2)  \ ,  \cr
 & & \cr
%%%%%%%%%
 \delta M_{12}^b & = &
{3 m_b^4 \lambda x A_t \Delta_{{\tilde b}_1} \Delta_{{\tilde b}_2}
\over 8 \pi^2 v^2 \cos^2 \beta}
{g(m_{{\tilde b}_1}^2, \ m_{{\tilde b}_2}^2)
\over (m_{{\tilde b}_2}^2 - m_{{\tilde b}_1}^2)^2}
- {3 m_b^2 \lambda x A_t \cos \phi \over 16 \pi^2 v^2 \cos^2 \beta}
f(m_{{\tilde b}_1}^2, \ m_{{\tilde b}_2}^2)  \cr
 & &\mbox{}
+ {3 m_b^4 \lambda x \Delta_{{\tilde b}_1} \over 8 \pi^2 v^2 \cos^2 \beta}
{\log (m_{{\tilde b}_2}^2 / m_{{\tilde b}_1}^2)
 \over (m_{{\tilde b}_2}^2 - m_{{\tilde b}_1}^2)} \cr
 & & \cr
%%%%%%%%%
 \delta M_{13}^b & = & \mbox{}
- {3 m_b^4 \lambda x A_t^2 \sin \phi \Delta_{{\tilde b}_2}
\over 8 \pi^2 v^2 \cos^3 \beta}
{g(m_{\tilde{b}_1}^2, \ m_{\tilde{b}_2}^2) \over
(m_{\tilde{b}_2}^2 - m_{\tilde{b}_1}^2)^2 }
+ {3 m_b^2 \lambda x A_t \sin \phi \over 16 \pi^2 v^2 \cos \beta}
f(m_{\tilde{b}_1}^2, \ m_{\tilde{b}_2}^2) \cr
 & &\mbox{}
- {3 m_b^4 \lambda x A_t \sin \phi \over 8 \pi^2 v^2 \cos^3 \beta}
{\log (m_{\tilde{b}_2}^2 / m_{\tilde{b}_1}^2) \over
(m_{\tilde{b}_2}^2 - m_{\tilde{b}_1}^2)}  \ , \cr
 & & \cr
%%%%%%%%%
 \delta M_{14}^b & = &
{3 m_b^4 \lambda A_t \Delta_{{\tilde b}_1} \Delta_{{\tilde b}_2}
\over 8 \pi^2 v \cos \beta \cot \beta}
{g(m_{\tilde{b}_1}^2, \ m_{\tilde{b}_2}^2) \over
(m_{\tilde{b}_2}^2 - m_{\tilde{b}_1}^2)^2}
- {3 m_b^2 \lambda A_t \cos \phi \over 16 \pi^2 v \cos \beta \cot \beta}
f(m_{\tilde{b}_1}^2, \ m_{\tilde{b}_2}^2) \cr
 & &\mbox{} + {3 m_b^4 \lambda \Delta_{{\tilde b}_1}
\over 8 \pi^2 v \cos \beta \cot \beta}
{\log (m_{\tilde{b}_2}^2 / m_{\tilde{b}_1}^2)
\over (m_{\tilde{b}_2}^2 - m_{\tilde{b}_1}^2) }  , \cr
 & & \cr
%%%%%%%%%
 \delta M_{15}^b & = & \mbox{}
- {3 m_b^4 \lambda A_t^2 \sin \phi \Delta_{{\tilde b}_2}
\over 8 \pi^2 v \cos \beta \cot \beta}
{g(m_{\tilde{b}_1}^2, \ m_{\tilde{b}_2}^2) \over
(m_{\tilde{b}_2}^2 - m_{\tilde{b}_1}^2)^2 }
+ {3 m_b^2 \lambda A_t \sin \phi \over 16 \pi^2 v \cos \beta \cot \beta}
f(m_{\tilde{b}_1}^2, \ m_{\tilde{b}_2}^2) \cr
 & &\mbox{}
- {3 m_b^4 \lambda A_t \sin \phi \over 8 \pi^2 v \cos \beta \cot \beta}
{ \log (m_{\tilde{b}_2}^2 / m_{\tilde{b}_1}^2) \over
(m_{\tilde{b}_2}^2 - m_{\tilde{b}_1}^2)}  \ , \cr
 & & \cr
%%%%%%%%%
 \delta M_{23}^b & = & \mbox{}
- {3 m_b^4 \lambda^2 x^2 A_t \sin \phi \Delta_{{\tilde b}_1}
\over 8 \pi^2 v^2 \cos^3 \beta}
{g(m_{\tilde{b}_1}^2, \ m_{\tilde{b}_2}^2) \over
(m_{\tilde{b}_2}^2 - m_{\tilde{b}_1}^2)^2 }
+ {3 m_b^2 \lambda x A_t \sin \phi \over 16 \pi^2 v^2 \cos \beta \cot \beta}
f(m_{\tilde{b}_1}^2, \ m_{\tilde{b}_2}^2) \ , \cr
 & & \cr
%%%%%%%%%
 \delta M_{24}^b & = &
{3 m_b^4 \lambda^2 x \Delta_{{\tilde b}_1}^2
\over 8 \pi^2 v \cos \beta \cot \beta}
{g(m_{\tilde{b}_1}^2, \ m_{\tilde{b}_2}^2)
\over (m_{\tilde{b}_2}^2 - m_{\tilde{b}_1}^2)^2 }
- {3 m_b^2 \lambda (\Delta_{{\tilde b}_1} + \lambda x \tan \beta)
\over 16 \pi^2 v \cos \beta}
f(m_{\tilde{b}_1}^2, \ m_{\tilde{b}_2}^2)  , \cr
 & & \cr
%%%%%%%%
 \delta M_{25}^b & = &\mbox{}
- {3 m_b^4 \lambda^2 x A_t \sin \phi \Delta_{{\tilde b}_1}
\over 8 \pi^2 v \cos \beta \cot \beta}
{g(m_{\tilde{b}_1}^2, \ m_{\tilde{b}_2}^2) \over
(m_{\tilde{b}_2}^2 - m_{\tilde{b}_1}^2)^2 }
+ {3 m_b^2 \lambda A_t \sin \phi \over 16 \pi^2 v \cos \beta \cot \beta}
f(m_{\tilde{b}_1}^2, \ m_{\tilde{b}_2}^2) \ , \cr
 & & \cr
%%%%%%%%%
 \delta M_{34}^b & = & \mbox{}
- {3 m_b^4 \lambda^2 x A_t \sin \phi \Delta_{{\tilde b}_1}
\over 8 \pi^2 v \cos^2 \beta \cot \beta}
{g(m_{\tilde{b}_1}^2, \ m_{\tilde{b}_2}^2)
\over (m_{\tilde{b}_2}^2 - m_{\tilde{b}_1}^2)^2 }
+ {3 m_b^2 \lambda A_t \sin \phi \over 16 \pi^2 v \cos^2 \beta}
f(m_{\tilde{b}_1}^2, \ m_{\tilde{b}_2}^2)  \ , \cr
 & & \cr
%%%%%%%%%
 \delta M_{35}^b & = & \mbox{}
{3 m_b^4 \lambda^2 x A_t^2 \sin^2 \phi
\over 8 \pi^2 v \cos^2 \beta \cot \beta}
{g(m_{\tilde{b}_1}^2, \ m_{\tilde{b}_2}^2)
\over (m_{\tilde{b}_2}^2 - m_{\tilde{b}_1}^2)^2 }
+ {3 m_b^2 \lambda A_t \cos \phi \over 16 \pi^2 v \cos^2 \beta}
f(m_{\tilde{b}_1}^2, \ m_{\tilde{b}_2}^2)  \ , \cr
 & & \cr
%%%%%%%%%
 \delta M_{45}^b & = & \mbox{}
- {3 m_b^4 \lambda^2 A_t \sin \phi \Delta_{{\tilde b}_1}
\over 8 \pi^2 \cot^2 \beta}
{ g(m_{\tilde{b}_1}^2, \ m_{\tilde{b}_2}^2)
\over (m_{\tilde{b}_2}^2 - m_{\tilde{b}_1}^2)^2 } \ .
\end{eqnarray}
%*****************************************************************************

In the above expressions, we have 
\begin{eqnarray}
 \Delta_{{\tilde t}_1} &=& A_t \cos \phi + \lambda x \cot \beta  \  , \cr
 \Delta_{{\tilde t}_2} & = & A_t + \lambda x \cot \beta \cos \phi \ ,
\end{eqnarray}
in $\delta M_{ij}^t$ and
\begin{eqnarray}
 \Delta_{{\tilde b}_1} &=& A_t \cos \phi + \lambda x \tan \beta  \  , \cr
 \Delta_{{\tilde b}_2} & = & A_t + \lambda x \tan \beta \cos \phi \ ,
\end{eqnarray}
in $\delta M_{ij}^b$, and we have the dimensionless functions $f$ and $g$ as 
\begin{eqnarray}
 f(m_1^2, \ m_2^2) & = & {1 \over (m_2^2-m_1^2)}
\left[  m_1^2 \log {m_1^2 \over \Lambda^2} -m_2^2
\log {m_2^2 \over \Lambda^2} \right] + 1 \ , \cr
 & & \cr
 g(m_1^2,m_2^2) & = & {m_2^2 + m_1^2 \over m_1^2 - m_2^2}
        \log {m_2^2 \over m_1^2} + 2 \ .
\end{eqnarray}

Note that $M_{13}^0$ and $M_{23}^0$ are zero in $M_{ij}^0$ at the tree level.
The former comes from the coupling between  $S_1$ and $A$ in the neutral Higgs potential, whereas the latter from the coupling between $S_2$ and $A$.
On the other hand, $M_{15}^0$, $M_{25}^0$, $M_{34}^0$, and $M_{45}^0$ are all non-zero. 
This implies that the scalar-pseudoscalar mixings at tree level may not occur between two Higgs doublets $H_1$ and $H_2$.
They can only occur either between the Higgs doublets and the singlet or in the cubic term of the Higgs singlet itself.
At the 1-loop level, the scalar-pseudoscalar mixings between two Higgs doublets are generated by the radiative corrections due to the quark and scalar quark of the third generation.
These scalar-pseudoscalar mixings are eventually responsible for the explicit CP violation in the radiatively corrected Higgs sector of the NMSSM.

Every element of the mass matrix which describes the scalar-pseudoscalar mixings among the neutral Higgs boson masses are nonzero and proportional to $\sin \phi$ at the 1-loop level.
Thus, the CP symmetry would be conserved in the Higgs sector if $\phi$ is zero [16]. 
The maximal CP violation occurs when $\sin \phi = 1$.  

The five physical neutral Higgs bosons are defined as the mass eigenstates, obtained by diagonalizing the mass matrix at 1-loop level, by the help of an orthogonal transformation matrix. 
The elements of the orthogonal transformation matrix, $O_{ij}$ ($i, j$=1 to 5), determine the couplings of the physical neutral Higgs bosons to the other states in the NMSSM.
Let us denote the five physical neutral Higgs bosons as $h_i$ ($i$=1 to 5). 
We assume that the mass of $h_i$ is smaller than $h_j$ if $i<j$.  

Note that when the vacuum expectation value (VEV) of the neutral Higgs singlet, $x$, is very large, one can derive the upper bound on the mass of the lightest neutral Higgs boson at tree level as [17]
\begin{equation}
(m_{h_1, {\rm max}}^0)^2 = m_Z^2 + (\lambda^2 v^2 - m_Z^2) \sin^2 2 \beta \ ,
\end{equation}
which is expressed in terms of only two parameters of the NMSSM: $\lambda$ and $\tan \beta$. 
None of $A_{\lambda}$, $k$, $A_k$, and $x$ does enter in the expression. 
In particular, $m_{h_1, {\rm max}}^0$ does not contain the CP violating phase $\phi$.

At 1-loop level the radiative corrections render the above expression.
Assuming both explicit CP violation and large $x$, the upper bound on the radiatively corrected mass of the lightest neutral Higgs boson is given as
\begin{eqnarray}
(m_{h_1, {\rm max}}^1)^2 & = & (m_{h_1}^0)^2 + {3 m_t^4 \over 8 \pi^2 v^2}
{(\lambda x \cot \beta \Delta_{{\tilde t}_1} + A_t \Delta_{{\tilde t}_2})^2 
\over (m_{{\tilde t}_2}^2 - m_{{\tilde t}_1}^2)^2}        
g(m_{{\tilde t}_1}^2, \ m_{{\tilde t}_2}^2) \cr
&     & \mbox{} + {3 m_t^4 \over 4 \pi^2 v^2}
{(\lambda x \cot \beta \Delta_{{\tilde t}_1} + A_t \Delta_{{\tilde t}_2}) 
\over (m_{{\tilde t}_2}^2 - m_{{\tilde t}_1}^2)}        
\log \left({m_{{\tilde t}_2}^2 \over m_{{\tilde t}_1}^2}\right) 
+ {3 m_t^4 \over 8 \pi^2 v^2}
\log({m_{{\tilde t}_1}^2 m_{{\tilde t}_2}^2 \over m_t^4}) \cr
& &\mbox{} + {3 m_b^4 \over 8 \pi^2 v^2}
{(\lambda x \tan \beta \Delta_{{\tilde b}_1} + A_t \Delta_{{\tilde b}_2})^2 
\over (m_{{\tilde b}_2}^2 - m_{{\tilde b}_1}^2)^2}        
g(m_{{\tilde b}_1}^2, \ m_{{\tilde b}_2}^2) \cr
&     & \mbox{} + {3 m_b^4 \over 4 \pi^2 v^2}
{(\lambda x \tan \beta \Delta_{{\tilde b}_1} + A_t \Delta_{{\tilde b}_2}) 
\over (m_{{\tilde b}_2}^2 - m_{{\tilde b}_1}^2)}        
\log \left({m_{{\tilde b}_2}^2 \over m_{{\tilde b}_1}^2}\right) 
+ {3 m_b^4 \over 8 \pi^2 v^2}
\log({m_{{\tilde b}_1}^2 m_{{\tilde b}_2}^2 \over m_b^4})    \ .
\end{eqnarray}
Note here that the upper bound is independent of the renormalization scale $\Lambda$.
If $\phi = 0$, there would be no CP violation, and taking $\cos \phi = 1$ in the above expression for $m_{h_1, {\rm max}}^1$ would reduce to the one obtained 
with CP conservation. 
It is clear that the upper bound on the lightest neutral Higgs boson mass decreases as the CP violating phase $\phi$ goes from 0 to $\pi$.

We focus on the charged Higgs sector. 
In the NMSSM the CP violation cannot occur in an explicit manner in the charged Higgs sector because the pair of charged Higgs bosons are CP even. 
In a unitary gauge, the charged Higgs boson mass is obtained by differentiating the Higgs potential twice with respect to the charged Higgs field. 
At the tree level the charged Higgs boson mass in the NMSSM is given as 
\begin{equation}
 (m_C^0)^2 = m_W^2 - \lambda^2 v^2 + \lambda (A_{\lambda} + k x \cos \phi)
{x (1 + \tan^2 \beta) \over \tan \beta} \ .
\end{equation}
One can find that the relative sizes of the second and the third term would make $m_C^0$ at the tree level either heavier or lighter than the $W$ boson.
Note that the dependence of $m_C^0$ on the CP violating phase $\phi$. 
If $A_{\lambda}$ be of the order of 1 TeV, $m_C^0$ would depend weakly on $\phi$, and if $A_{\lambda}$ be as small as O(0.1 TeV), it would strongly depend on $\phi$. 
Moreover, unlike the neutral Higgs sector, the charged Higgs boson mass at the tree level depends on all the six parameters: $\tan \beta$, $\lambda$, $A_{\lambda}$, $k$, $x$, and $\phi$.
The dependence of $m_C^0$ on $\lambda$ is not clearly visible in above expressions. 
However, it is easy to see that the mass of the charged Higgs boson would increase if any one of $\tan \beta$ $(> 1)$, $A_{\lambda}$, $k$, $x$, and $\cos \phi$ increases.
 
The radiative corrections to the charged Higgs boson mass contain the contributions from the scalar-quark loops which depend on the scalar-quark masses of the third generation. 
Thus, one has to know the scalar-quark masses in advance.
On the other hand, the scalar-quark masses depend on the charged Higgs field: The scalar-quark masses are given as the eigenvalues of a $4\times 4$ Hermitian mass matrix ${\cal M}_{\tilde t}$ whose elements are coupled to the charged Higgs field.
It is difficult to obtain an analytical expression for the scalar-quark masses in terms of the charged Higgs field.
One can calculate the first derivative of the squared masses of the scalar quark with respect to the charged Higgs field. 
By replacing the result with VEVs, one can find a trivial fact.
The only nontrivial part of the calculation is the evaluation of the second derivatives of the field dependent scalar-quark masses with respect to the charged Higgs field.

The radiatively corrected mass of the charged Higgs boson is given as follows [18]: 
\begin{eqnarray}
m_C^2 & = & (m_C^0)^2 + {3 m_t^2 m_b^2 \over 2 \pi^2 v^2 \sin^2 2 \beta}
        f(m_t^2, \ m_b^2)   \cr
 & &\mbox{} - {3 \over \pi^2 v^2 \sin^4  2 \beta}
\sum^4_{i = 1} m_{{\tilde q}_i}^2 \left(\log{m_{{\tilde q}_i}^2 \over \Lambda^2} - 1 \right)
{B m_{{\tilde q}_i}^4 + C m_{{\tilde q}_i}^2 + D \over \prod\limits_{j \not= i} (m_{{\tilde q}_i}^2 - m_{{\tilde q}_j}^2)} \ ,
\end{eqnarray}
where $m_{{\tilde q}_i}^2$ are the scalar-quark masses for the third generation. 
The expressions for $B$, $C$, and $D$ are given as
%**************************  charged Higgs *****************************
\begin{eqnarray}
 B & = & B_{\phi}
+ \cos \beta \sin \beta
\{A_t \lambda x (m_b^2 \sin^2 \beta + m_t^2 \cos^2 \beta)
- m_b^2 m_t^2  \sin 2 \beta \}  \ , \cr
 C & = & C_{\phi}
+ \lambda^2 x^2 (m_b^4 \sin^4 \beta + m_t^4 \cos^4 \beta)
- \lambda x A_t m_t^2 (2 m_b^2 + m_Q^2 + m_T^2) \cos^3 \beta \sin \beta \cr
 & &\mbox{}
- \lambda x A_t m_b^2 (2 m_t^2 + m_Q^2 + m_T^2) \cos \beta \sin^3 \beta \cr
 & &\mbox{}
+ 2 m_b^2 m_t^2 (m_b^2 + m_t^2 - 2 A_t^2 + m_Q^2 + m_T^2 + \lambda^2 x^2) \cos^2 \beta \sin^2 \beta   \ , \cr
 D & = &\mbox{} D_{\phi}
- \lambda^2 x^2 m_t^4 (m_b^2 + m_T^2) \cos^4 \beta
- \lambda^2 x^2 m_b^4 (m_t^2 + m_T^2) \sin^4 \beta \cr
 & &\mbox{}
- \lambda x A_t m_t^2 (m_b^2 (A_t^2 + 2 m_t^2 + \lambda^2 x^2)
- (m_b^2 + m_Q^2)(m_b^2 + m_T^2)) \cos^3 \beta \sin \beta  \cr
 & &\mbox{}
- \lambda x A_t m_b^2 (m_t^2 (A_t^2 + 2 m_b^2 + \lambda^2 x^2)
- (m_t^2 + m_Q^2)(m_t^2 + m_T^2)) \cos \beta \sin^3 \beta  \cr
 & &\mbox{} - m_b^2 m_t^2 \{
2 m_b^2 m_t^2 + m_T^4 - A_t^2 (m_b^2 + 4 m_Q^2 + m_t^2 + 2 m_T^2)
+ (A_t^2 + m_Q^2 + \lambda^2 x^2)^2  \cr
 & &\mbox{} + (m_b^2 + m_t^2)(m_Q^2 + m_T^2 + 2 \lambda^2 x^2)
\} \cos^2 \beta \sin^2 \beta  \ , 
\end{eqnarray}
%********************************************************************
where
\begin{eqnarray}
 B_{\phi} & = & - A_t \lambda x (m_b^2 \sin^2 \beta + m_t^2 \cos^2 \beta)
 \sin 2 \beta \sin^2 {\phi \over 2} \ , \cr
 & & \cr
 C_{\phi} & = & {A_t \over 2}
[4 m_b^2 m_t^2 \lambda x + \lambda x (m_b^2 + m_t^2)(m_Q^2 + m_T^2)
+ \lambda x (m_t^2 - m_b^2)(m_Q^2 + m_T^2) \cos 2 \beta  \cr
 & &\mbox{}
+ 4 A_t m_t^2 (3 m_b^2 + m_t^2 + (m_t^2 - m_b^2) \cos 2 \beta) (1 + \cos \phi) \cos^3 \beta \sin \beta
] \sin 2 \beta \sin^2 {\phi \over 2} \ , \cr
 & & \cr
 D_{\phi} & = & - {A_t \over 8} (
2 \lambda x m_t^2 (2 m_Q^2 m_T^2 - 3 A_t^2 m_b^2 - 2 m_b^4) \cr
 & &\mbox{}
+ 4 m_b^2 \lambda x \{m_Q^2 (2 m_t^2 + m_T^2) - m_t^2 (m_t^2 - 2 m_T^2 + 2 \lambda^2 x^2) \} \cr
 & &\mbox{}
+ \lambda x \{A_t^2 m_b^2 m_t^2 + 4 (m_b^2 - m_t^2)(3 m_b^2 m_t^2 - m_Q^2 m_T^2) \} \cos 2 \beta \cr
 & &\mbox{}
+ A_t m_t^2 \{32 \lambda x A_t m_b^2 \cos^4 \beta \sin^2 \beta \cos 2 \phi
- 2 \lambda x A_t m_b^2 \cos 4 \beta - \lambda x A_t m_b^2 \cos 6 \beta \cr
 & &\mbox{}
+ (3 m_b^4 - 3 A_t^2 m_b^2 + 8 m_b^2 m_Q^2 + 13 m_b^2 m_t^2 + 3 m_b^2 m_T^2
+ 5 m_t^2 m_T^2 - 5 \lambda^2 x^2 m_b^2 ) \sin 2 \beta \cr
 & &\mbox{} + 2 [
m_t^2 (7 m_b^2 + 3 m_T^2) + m_b^2 (4 m_Q^2 + m_T^2 - A_t^2 + m_b^2 - 3 \lambda^2 x^2) \cr
 & &\mbox{} + 4 \{m_t^2 (2 m_b^2 + m_T^2) + m_b^2 (m_Q^2 + \lambda^2 x^2) \}
\cos 2 \beta \cr
 & &\mbox{}
+ \{m_b^2 (A_t^2 - \lambda^2 x^2) - (m_b^2 - m_t^2)(m_b^2 + m_T^2)
\} \cos 4 \beta \cr
 & &\mbox{}
+ 16 \lambda x A_t m_b^2 \cos^3 \beta \sin \beta
] \sin 2 \beta \cos \phi \cr
 & &\mbox{}
+ 4 (m_b^2 m_Q^2 + 2 m_b^2 m_t^2 + m_t^2 m_T^2 + \lambda^2 x^2 m_b^2
) \sin 4 \beta\cr
 & &\mbox{}
+ (A_t^2 m_b^2 - (m_b^2 - m_t^2)(m_b^2 + m_T^2) - \lambda^2 x^2 m_b^2
) \sin 6 \beta
\}) \sin 2 \beta \sin^2 {\phi \over 2}  \ .
\end{eqnarray}
%********************************************************************

In the formula for the charged Higgs boson mass [Eq. (15)], the second and the third terms represent the radiative corrections: The second term comes from the quark loops of the third generation, and the third one from the scalar-quark loops of the third generation.
As mentioned earlier, if the scalar-quark masses are degenerate in the 1-loop effective potential, radiative corrections to the charged Higgs boson mass would vanish and both the second and the third terms would be absent.

%*****************************************************************
\section{PHENOMENOLOGICAL ANALYSIS}
%*****************************************************************

We consider the neutral Higgs production in $e^+ e^-$ collisions in order to investigate the effects of the explicit CP violation in the NMSSM, and to impose phenomenological constraints on the parameter space using the LEP2 data. 
For the center of mass energy at LEP2, the main contributions to the production cross section for the neutral Higgs boson come from the Higgs-strahlung process and the Higgs-pair production process.
The Higgs-strahlung process is  
\[
\begin{array}{ccccc}
 ({\rm i}) & e^+ e^- & \rightarrow Z^*  & \rightarrow & Z h_i 
\ (i=1 \ {\rm to} \ 5) \ ,
\end{array}
\]
where the $Z$ boson that is emitted together with the neutral Higgs boson in the final state is real for the center of mass energy at LEP2.
When the next-to-lightest neutral Higgs boson mass is relatively small, the cross section of the Higgs-pair production process is comparable to that of the Higgs-strahlung one.
The Higgs-pair production process is 
\[
\begin{array}{ccccc}
 ({\rm ii}) & e^+ e^- & \rightarrow Z^* & \rightarrow & h_i h_j 
\ {\rm for} \ (i, j=1 \ {\rm to} \ 5, i \neq j) \ .
\end{array}
\]
The relevant couplings for the above two processes are
\begin{eqnarray}
 G_{Z Z h_i}^2 & = &\mbox{} {g_2^2 m_Z^2 \over 4 \cos^2 \theta_W} (\cos \beta O_{i 1} + \sin \beta O_{i 2})^2 \ , \cr
 G_{Z h_i h_j}^2 & = & {g_2^2 \over 4 \cos^2 \theta_W}
[ \{O_{j 3} (\cos \beta O_{i 2} - \sin\beta O_{i 1})
- O_{i 3} (\cos \beta O_{j 2} - \sin\beta O_{j 1}) \}^2 \cr
 & &\mbox{} + \{O_{i 4} O_{j 4} + O_{i 5} O_{j 5} \}^2 ] \ ,
\end{eqnarray}
where $\theta_W$ is the Weinberg mixing angle, and $O_{ij}$ is the elements of the orthogonal transformation matrix that diagonalizes the mass matrix of the neutral Higgs bosons at 1-loop level.
Thus, both couplings include radiative corrections and the effects of explicit CP violation.

For numerical analysis, we input the values of relevant parameters in the NMSSM into our calculations. 
The masses of top quark [19] and bottom quark are fixed as 175 GeV and 4 GeV, respectively.
We assume that the mass of the lighter scalar-quark of the third generation is larger than the top quark mass.
Then, we fix the renormalization scale in the effective potential at 500 GeV.
We allow the remaining free parameters to vary within the following ranges: $2 \leqslant \tan \beta \leqslant 40$, $0 < \lambda \leqslant 0.87$, and $0 < k \leqslant 0.63$.
We assume that $A_{\lambda}$, $A_k$, $x$, $m_Q$, and $m_T$ may vary between 0 and 1000 GeV, while $A_t$ between 0 and 2000 GeV. 
The CP violating phase $\phi$ is assumed to take any value between 0 and $\pi$.

With those input values and ranges, we plot the contours of $m_{h_1}$ in Fig. 1 as functions of $A_{\lambda}$ and $A_k$, for $\tan \beta = 2$, $\lambda = 0.12$, $k$ = 0.04, $x = m_Q = 1000$ GeV, and $m_T = A_t = 500$ GeV. 
We take the CP violation phase to be $\phi = \pi/2$, the maximal CP violation.
In the same figure, the cross section $\sigma_t$ for scalar Higgs production in $e^+ e^-$ collisions at $\sqrt{s} = 200$ GeV is also shown, which is the sum of the cross section for Higgs-pair production process and that for Higgs-strahlung process.
The hatched regions in the figure are where $\sigma_t$ is larger than 0.1 pb and where $m_{h_1} \geqslant 0$.
If the values of the parameters of the NMSSM are such that they are represented
by points in the hatched regions, it is likely that a scalar Higgs boson might be detected at LEP2 with $\sqrt{s} = 200$ GeV, if the discovery limit of LEP2 is assumed to be about 0.1 pb.
In other words, taking the results of the Higgs search at LEP2 into account, one should exclude the hatched regions in Fig. 1.

On the other hand, the remaining region within the contour of $m_{h_1} = 0$ is where $\sigma_t < 0.1$ pb. 
Therefore, if the parameters take the set of values that are represented by points in that region, a scalar Higgs boson might exist without being detected at LEP2 with $\sqrt{s}$=200 GeV. 
Consequently, the results of the Higgs search at LEP2 at $\sqrt{s}$=200 GeV would not be able to put any constraints on $m_{h_1}$. 
The LEP2 data does not exclude the possibility of the existence of a massless Higgs boson in the NMSSM with explicit CP violation.

In Fig. 2(a), we plot the charged Higgs mass as function of the CP violation phase in order to see the effects of the explicit CP violation in the NMSSM, for two different values of $A_{\lambda}$: 200 and 400 GeV. 
The remaining parameters are set as $A_k = 50$ GeV, $\tan \beta = 2$, 
$\lambda = 0.12$, $k = 0.04$, $x = m_Q = 1000$ GeV, and $m_T = A_t =500$ GeV.
Those parameter values are consistent with the phenomenological constraints of LEP2 at $\sqrt{s} = 200$ GeV.
The two solid curves correspond to the tree-level masses of the charged Higgs boson, and the two dashed curves to the radiatively corrected ones. 
In both case, the upper one for $A_{\lambda} = 200$ GeV and the lower one to $A_{\lambda} = 400$ GeV. 
We find that the tree-level mass of the charged Higgs boson is roughly stable against the variation of the CP violation phase while the 1-loop level one decreases slightly as as $\phi$ goes from zero to $\pi$.
From the figure, it is clearly shown that the radiative corrections to the charged Higgs boson mass are negative for $0 \leqslant \phi \leqslant \pi$, and that the amount of radiative corrections decrease as the CP violation phase goes from zero to $\pi$.
Another point to note is that for the whole range of $\phi$ from 0 to $\pi$, the mass of the charged Higgs boson for $A_{\lambda} = 400$ GeV is found to be larger than that for $A_{\lambda} = 200$ GeV. 
This behavior does not change if the 1-loop contributions are included. 
Note that similar patterns are exhibited in the MSSM.

In the lower part of the same figure, there are a dotted curve and a dash-dotted one. 
They are the plots of the radiatively corrected masses of the lightest neutral scalar Higgs boson, $m_{h_1}$, as function of the CP violation phase, for the same set of parameter values.
The dotted curve corresponds to $m_{h_1}$ for $A_{\lambda}$=200 GeV and the dot-dashed curve for $A_{\lambda}$=400 GeV.
Unlike the charged Higgs mass, one can find that there occurs a crossover between the two curves as $\phi$ moves from zero to $\pi$: The mass for $A_{\lambda}$=200 GeV decreases as $\phi$ increases, whereas the one for $A_{\lambda}$=400 GeV increases as $\phi$ increases.

We repeat the same calcuations as Fig. 2(a) by changing the value of $\tan \beta$ from 2 to 10. 
The result is plotted in Fig. 2(b). 
In both figures, the values of the other parameters are same, and the figure captions are also the same.
In Fig. 2(b), one can see that the behavior of the mass of the charged Higgs boson for $\tan\beta = 10$ is quite similar to the one for $\tan \beta = 2$,
except that the mass of the charged Higgs boson for $\tan\beta$ = 10 is always larger by about 270 GeV than the one for $\tan \beta = 2$, regardless of the value of $\phi$. 
From Figs. 2(a) and 2(b), we may infer that independently of the value of $\tan \beta$, the radiative corrections to the charged Higgs boson mass contribute more negatively as the CP violation phase increases.

However, the behavior of $m_{h_1}$ for $\tan\beta = 10$ is different from the one for $\tan \beta = 2$. 
As we noted earlier, there is a crossover for $\tan\beta = 2$ between $m_{h_1}$ for $A_{\lambda} = 200$ GeV and the one for $A_{\lambda} = 400$ GeV as $\phi$ moves from zero to $\pi$. 
For $\tan \beta = 10$, there occurs no crossover at all between the curves for $A_{\lambda} = 200$ and 400 GeV. 
Figure 2(b) shows that $m_{h_1}$ remains roughly constant as $\phi$ varies, for $\tan \beta = 10$. 

We perform the calculations repeatedly by changing the values of the relevant parameters to examine if peculiar behaviors might be found.
For wide ranges of parameter values, no such peculiarity is found.
The detailed numerical results are similiar to the behaviors shown in Figs. 2.
In the presence of explicit CP violation, the radiative corrections to the mass of the charged Higgs boson is negative for $0 \leqslant \phi \leqslant \pi$, whereas those to the mass of the lightest neutral Higgs boson may be either negative or positive contributions, depending on the particular set of parameter values.

We turn to the Higgs production in $e^+e^-$ collisions.
We feel that it is worthwhile to investigate whether the LEP2 data at $\sqrt{s} = 200$ GeV can put a phenomenological constraint on the values of the NMSSM parameters.
In Fig. 3, we plot the lower bounds on $x$ as function of $\tan \beta$ for two different values of $\phi$. 
The solid curve corresponds to $\phi = 0$ (no CP violation) and the dashed one to $\phi = \pi/2$ (maximal CP violation). 
All the other parameters are allowed to vary freely only if they satisfy the condition that $\sigma_t$ should be less than 0.1 pb, by using a random number generating function.

The lower region of each curve is where the cross section for the Higgs production in $e^+e^-$ collisions at $\sqrt{s} = 200$ GeV, $\sigma_t$, is larger than 0.1 pb, and thus should be excluded. 
We can say that $x$ may by no means be smaller than 16 GeV for any set of values for other relevant parameters of the NMSSM, if the LEP2 data at $\sqrt{s} = 200$ GeV is taken into account with a discovery limit of 0.1 pb. 
Thus, 16 GeV is a kind of universal lower bound on $x$ in the NMSSM with explicit CP violation.
Figure 3 also tells us that the lower bound on $x$ increases up to as much as 37 GeV as $\tan \beta$ decreases to 10 down from 40.
For $5 \leqslant \tan \beta \leqslant 10$, note that the lower bound on $x$ for $\phi = 0$ is larger than that for $\phi = \pi/2$. 

We extend the preceding analysis to other relevant parameters of the NMSSM to see whether the LEP2 data at $\sqrt{s} = 200$ GeV can also put similar constraints on them. 
The procedure begins by generating random numbers for the parameter values, for a fixed value of the CP violating phase.
For a partucluar set of parameter values, we calculate the masses of the neutral scalar Higgs bosons and check if the center of mass energy of the $e^+e^-$ collider, $\sqrt{s}$, is sufficient to produce the neutral Higgs boson via the aforementioned two processes. 
If the center of mass energy of the collider is larger than $E_{\rm T}$ = $m_Z + m_{h_i}$, then the Higgs-strahlung process is viable for the Higgs production.
If $\sqrt{s}$ is still larger than $m_{h_i} + m_{h_j}$, the Higgs-pair production process is also allowed. 

In this way we select the kinematically accpetable sets of the parameter values from $10^7$ points in the parameter space. 
The selected sets of the parameter values form a subspace within the whole parameter space of the NMSSM, which one may conveniently call as the kinematically accpetable parameter space of the NMSSM. 
The points in the kinematically acceptable space are further classified according to whether the cross section for the Higgs production they yield is smaller than 0.1 pb or not. 
If $\sigma_t$ is smaller than 0.1 pb, we finally choose the points. 
Otherwise, we discard them.  
The result is roughly an estimate of the fraction of the kinematically acceptable parameter space which can produce the neutral Higgs bosons without contradicting the LEP2 data at $\sqrt{s} = 200$ GeV.
We may call this subspace of the kinematically acceptable space as the LEP2-allowed parameter space of the NMSSM.

In order to see the effect of the CP phase $\phi$ on the result, we repeat the procedure for different values of $\phi$, each time by generating $10^7$ sets of the parameter values.
The results are listed in Table 1. 
We see that about 44\% of the kinematically allowed parameter space should be excluded.
Thus, the size of the LEP2-allowed parameter space is approximately one half of the kinematically acceptable parameter space.
Although the ratio of the excluded part in the case of the maximal CP violation ($\phi = \pi/2$) is larger than those of other values of $\phi$, the excluded fraction over the parameter space does not significantly depend on the CP phase $\phi$.
In other words, one may say that in the explicit CP violation scenario the data from LEP2 at $\sqrt{s} = 200$ GeV can investigate about 44\% for the kinematically acceptable NMSSM parameter space.
 
With the knowledge on the allowed values of the parameters at hand, we proceed to estimate the mass of the charged Higgs boson in more detail.
We generate randomly the values of the relevant parameters to select points in the LEP2-allowed space. 
In this way, $10^7$ points are selected.
For each point in the LEP2-allowed space, we calculate the mass of the charged Higgs boson, and plot it in the ($m_C$, $x$)-plane.
The results are shown in Fig. 4(a) for $\phi = 0$ (no CP violation), and in Fig. 4(b) for $\phi = \pi/2$ (maximal CP violation).
Each figure contains the solid, dashed, dotted, and dot-dashed curves; they correspond to four different values of $\tan\beta = 2$, 5, 10, and 40. 

Those curves are actually the boundaries of the plotted points in the ($m_C$, $x$)-plane. 
Since there are no points plotted in the lower regions of those curves, the curves are clearly the lower bounds on the mass of the charged Higgs boson in the NMSSM.
There are no curves for smaller values of $x$. 
This is because the smaller $x$ is excluded, as can be seen in Fig. 3, by the condition of $\sigma_t < 0.1$ pb, which is the assumed discovery limit for LEP2 at $\sqrt{s} = 200$ GeV.

In Fig. 4(a), we find that minimum values of the lower bound on the charged Higgs mass for $\tan \beta = 40$, 10, and 5 are about 100, 95, and 85 GeV, respectively. 
If $\tan \beta$ is as small as 2, the minimum of the lower bound on the charged Higgs mass might be almost zero. 
This implies that the NMSSM may even have a nearly massless charged Higgs boson, without contradicting the LEP2 data, if the relevant parameters have certain values.

The minimum value of the lower bound on $m_C$ seems to increase sharply as $\tan \beta$ increases. 
In addition, the minimum value for $\tan \beta$ rises rapidly as $x$ increases.
For $300 \leqslant x \leqslant 1000$ GeV, the charged Higgs boson might not have a mass smaller than about 110 GeV, regardless of the value of $\tan \beta$.
The lower bounds for $\tan \beta = 5$, 10 and 40 are smaller than that for $\tan \beta = 2$ in the range of $300 \leqslant x \leqslant 1000$ GeV.

The above result is slightly different from, but essentially compatible with, the result of previous analyses [20]. 
In case of no CP violation, the previous analyses have estimated the lower bound on the charged Higgs boson mass is about 70 GeV. 
This number has been obtained under the assumption that the values of the relevant parameters are such that the mass difference between top scalar quarks are maximal. 
We do not assume to impose such constraints on the parameter values.
The parameter values are randomly generated within the LEP2-allowed space.
Thus, the top scalar quarks may have either maximally different masses or degenerate masses. 
In this sense, we may say that our result is more general.

In Fig. 4(b), we repeat our analysis by substituting the value of $\phi = 0$ by $\pi/2$, to investigate the effect of maximal CP violation.
Note that, as we assume that at the tree level the CP phase is equal to that at the 1-loop level, the CP violation arise maximally both at the tree level and at the 1-loop level.
The behavior of the curves are basically the same as the case of $\phi = 0$.
we find that minimum values of the lower bound on the charged Higgs mass for $5 \leqslant \tan \beta \leqslant 40$ is about 110 GeV, and nearly stable if $x$ is larger than about 100 GeV. 
The minimum values for $\phi = \pi/2$ are larger than those for $\phi = 0$, by about 25, 15, and 5 GeV for $\tan \beta = 5$, 10, and 40, respectively.
These increases may be regarded as the effect of the CP violation.

If $\tan \beta$ is 2, and if $x$ is about 75 GeV, there is a narrow possibility
that the charged Higgs might be massless, without contradicting the LEP2 data, even if CP violation is maximal.
Then, the minimum value of the lower bound on $m_C$ seems to increase abruptly as $\tan \beta$ increases from 2 upward. 
For $300 \leqslant x \leqslant 1000$ GeV, the charged Higgs boson might not have a mass smaller than about 110 GeV, regardless of the value of $\tan \beta$.

%*****************************************************************
\section{CONCLUSIONS}
%*****************************************************************

In the next-to minimal supersymmetric standard model, we consider the explicit 
CP violation scenario in its Higgs sector and the subsequent phenomenology. 
We calculate the masses of both the charged Higgs boson and the neutral ones at 1-loop level, by taking into account the radiative corrections due to the quarks and scalar-quarks of the third generation within the explicit CP violation scenario.
The effective potential at the 1-loop level, as well as the one at the tree level, is shown to have the CP violation phase.
It would be absent if the top scalar-quarks are degenerate in mass.

In the whole parameter space of the NMSSM we establish a subspace with two constraints. 
The first constraint is that for the values of the relevant parameters in the subspace it is kinematically possible to produce neutral Higgs bosons, and the second one is that the cross section for their productions does not exceed 0.1 pb. 
In this way, we establish the LEP2-allowed parameter space of the NMSSM, as we assume the discovery limit of LEP2 at $\sqrt{s} = 200$ GeV.
The lower bound on $x$, the VEV of the neutral Higgs singlet, is found to be about 16 GeV. 
It is shown to increase to 40 GeV as $\tan \beta$ approaches 2. 

We find that the LEP2 data do not exclude the possibility of a massless neutral Higgs boson. 
For the charged Higgs boson, we find that the radiative corrections to its mass are negative.
For $\tan \beta = 2$, we find that for certain values of $x$ the lower bound on the radiatively corrected mass of the charged Higgs boson is as low as zero, whether CP is violated maximally or not at all.
The minimum of the lower bound on the charged Higgs boson mass increases as $\tan \beta$ increases.

For larger values of $\tan \beta$, the minimum value of the lower bound on the charged Higgs boson mass for maximal CP violation is found to be larger than the one for no CP violation. 
If $x$ takes a value between 300 GeV and 1000 GeV, the lower bound on the mass of the charged Higgs boson in the NMSSM is about 110 GeV, for $2 \leqslant \tan\beta \leqslant 40$, whether CP is violated maximally or not at all. 

%*****************************************************************
\vskip 0.3 in

\noindent
{\large {\bf ACKNOWLEDGMENTS}}

This work was supported by the grant of Post-Doc. Program, Kyungpook National
University (2000).
This research is partly supported through the Science Research Center Program by the Korea Science and Engineering Foundation.

%\vskip 0.3 in
%\vfil\eject

%******************************************************************

\vfil\eject

{\large {\bf TABLE CAPTION}}
%\vskip 0.3 in
%\noindent

%Table
\setcounter{table}{0}
\def\tablename{}{}%
\renewcommand\thetable{TABLE 1}
\begin{table}[h]
\caption{Numerical results of $\sigma_t$ at $\sqrt{s} = 200$ GeV for the parameter space of $0 < \tan \beta \leqslant 40$, $0 \lambda \leqslant 0.87$, $0 < k \leqslant 0.63$, 0 $< A_{\lambda}$, $A_k$, $x$, $m_Q$, $m_T$ $\leqslant$ 1000 GeV, 0 $< A_t \leqslant$ 200 GeV.
The number of $10^6$ points is kinematically allowed for the above parameter space at $\sqrt{s} = 200$ GeV.}
%\vskip 5pt
%\tabcolsep=7.9pt
\begin{center}
\begin{tabular}{c|c|c|c|c} \hline\hline
$\phi$         & 0      & $\pi$/4 & $\pi$/2  & 3 $\pi$/4  \\ 
\hline\hline
$\sigma_t \geqslant 0.1$ pb & 446668 & 446004  & 454740   &  432038    \\ 
\hline
$\sigma_t < 0.1$ pb  & 553332 & 553996  & 545260   &  567962    \\ 
\hline\hline
\end{tabular}
\end{center}
\end{table}

{\large {\bf FIGURE CAPTION}}
\vskip 0.3 in
\noindent
FIG. 1. : The contours of $m_{h_1}$ and $\sigma_t$ in $e^+ e^-$ collisions at $\sqrt{s} = 200$ GeV, as functions $A_{\lambda}$ and $A_k$, for $\phi = \pi/2$, $\tan \beta = 2$, $\lambda = 0.12$, $k = 0.04$, $x = m_Q = 1000$ GeV, and $m_T = A_t = 500$ GeV.

\vskip 0.2 in
\noindent
FIG. 2. (a) : The parameters are set as $A_k = 50$ GeV, $\tan \beta = 2$, $\lambda = 0.12$, $k = 0.04$, $x = m_Q = 1000$ GeV, and $m_T = A_t = 500$ GeV.
The dotted and dot-dashed curves are for $m_{h_1}$, as a function $\phi$, for $A_{\lambda} = 200$ and 400 GeV, respectively. 
The upper and lower solid (dashed) curves are for $m_C^0$ ($m_C$), as a function $\phi$, for $A_{\lambda} = 200$ and 400 GeV, respectively. 

\vskip 0.2 in
\noindent
FIG. 2. (b) : The same as Fig. 2(a), except for $\tan \beta = 10$.  
 
\vskip 0.2 in
\noindent
FIG. 3 : The lower bound on $x$ against $\tan \beta$ under the condition of $\sigma_t < 0.1$ pb for $\phi = 0$ and $\phi = \pi/2$. 
In this calculations, the remaining parameters are independently varied for $0 < \lambda \leqslant 0.87$, $0 < k \leqslant 0.63$, $0 < A_{\lambda}, A_k, m_Q, m_T \leqslant 1000$ GeV, and $0 < A_t \leqslant 2000$ GeV.

\vskip 0.2 in
\noindent
FIG. 4. (a) : The lower bound on $m_C$ under the condition of $\sigma_t < 0.1$ pb as a function of $x$, for $\tan \beta = 2$ (solid curve), 5 (dashed curve), 10 (dotted curve), and 40 (dot-dashed curve). In this calculation, the remaining parameters are independently varied for the same parameter space as Fig. 3.

\vskip 0.2 in
\noindent
FIG. 4. (b) : The same as Fig. 4(a) except for $\phi = \pi/2$.

\vfil\eject

%Figure
\setcounter{figure}{0}
\def\figurename{}{}%
% (FIG 1)
\renewcommand\thefigure{FIG. 1.}
\begin{figure}[t]
\epsfxsize=13cm
\hspace*{2.cm}
\epsffile{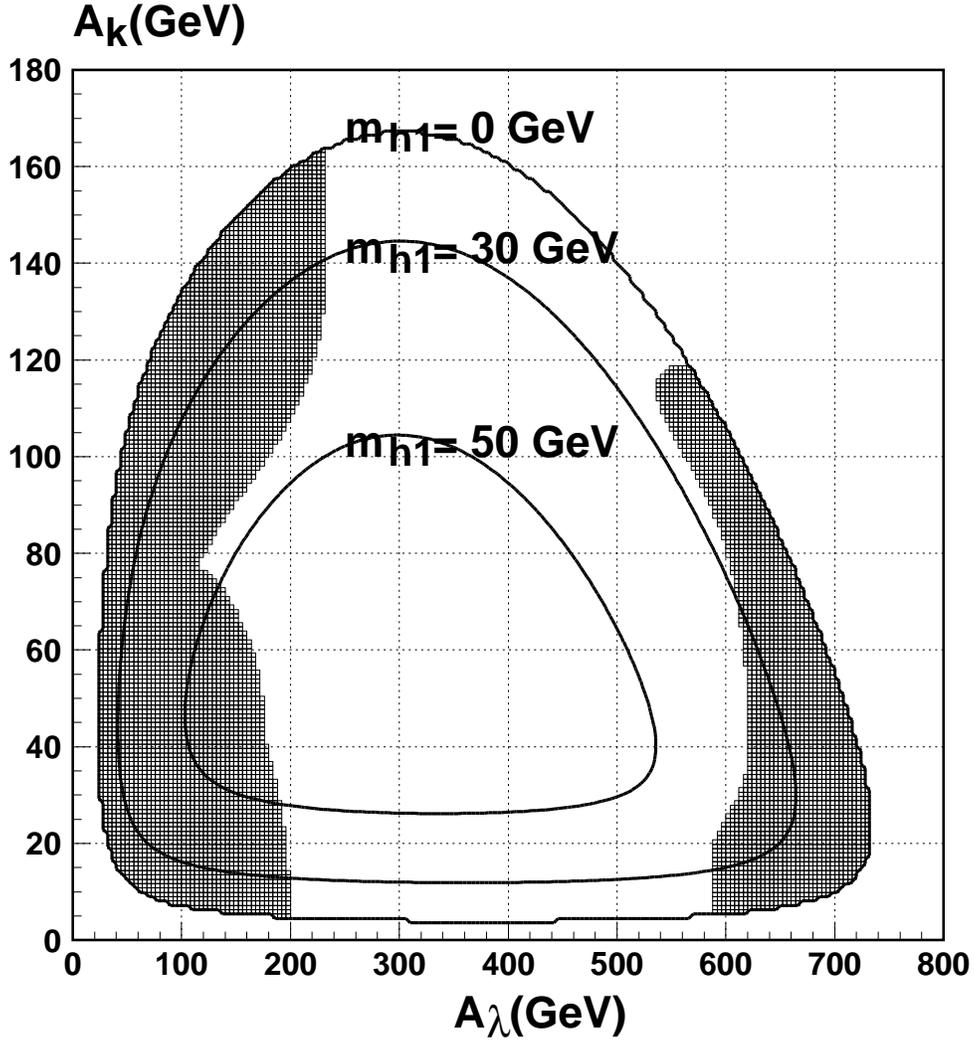}
\caption[plot]{The contours of $m_{h_1}$ and $\sigma_t$ in $e^+ e^-$ collisions at $\sqrt{s} = 200$ GeV, as functions $A_{\lambda}$ and $A_k$, for $\phi = \pi/2$, $\tan \beta = 2$, $\lambda = 0.12$, $k = 0.04$, $x = m_Q = 1000$ GeV, and $m_T = A_t = 500$ GeV.}
\end{figure}
% (FIG 2a)
\renewcommand\thefigure{FIG. 2. (a)}
\begin{figure}[t]
\epsfxsize=13cm
\hspace*{2.cm}
\epsffile{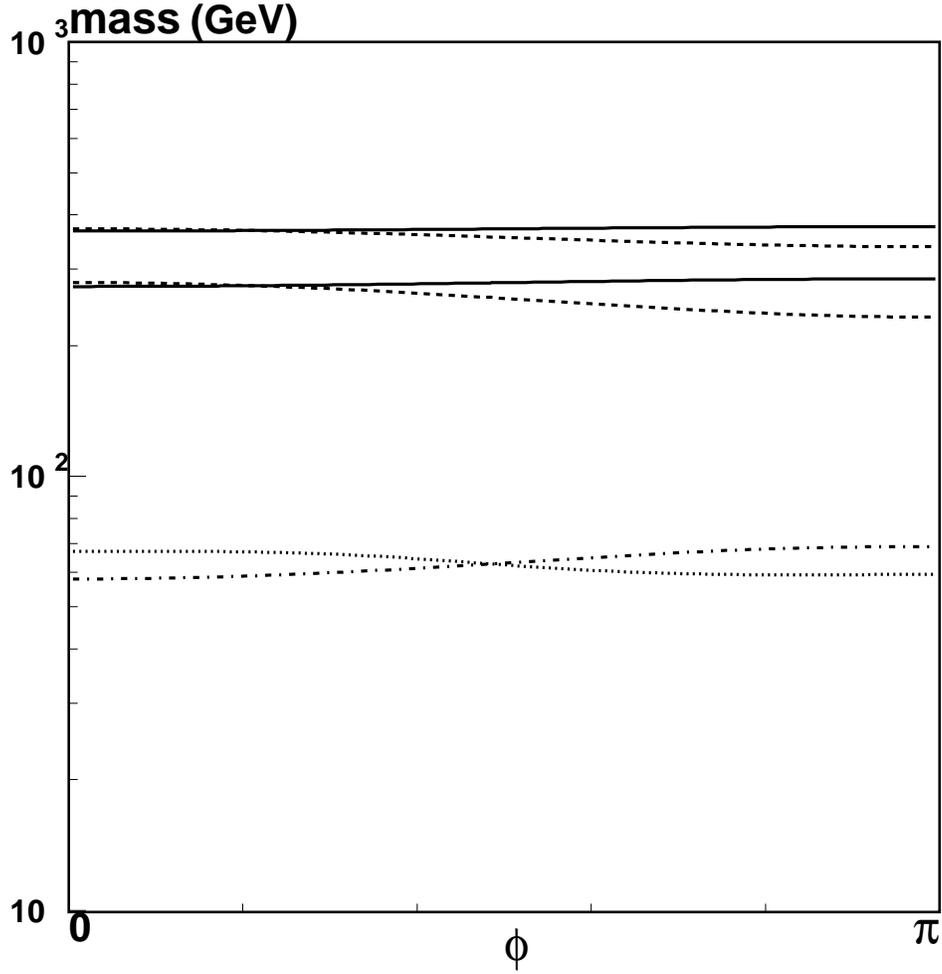}
\caption[plot]{The parameters are set as $A_k = 50$ GeV, $\tan \beta = 2$, $\lambda = 0.12$, $k = 0.04$, $x = m_Q = 1000$ GeV, and $m_T = A_t = 500$ GeV.
The dotted and dot-dashed curves are for $m_{h_1}$, as a function $\phi$, for $A_{\lambda} = 200$ and 400 GeV, respectively. 
The upper and lower solid (dashed) curves are for $m_C^0$ ($m_C$), as a function $\phi$, for $A_{\lambda} = 200$ and 400 GeV, respectively.}
\end{figure}
% (FIG 2b)
\renewcommand\thefigure{FIG. 2. (b)}
\begin{figure}[t]
\epsfxsize=13cm
\hspace*{2.cm}
\epsffile{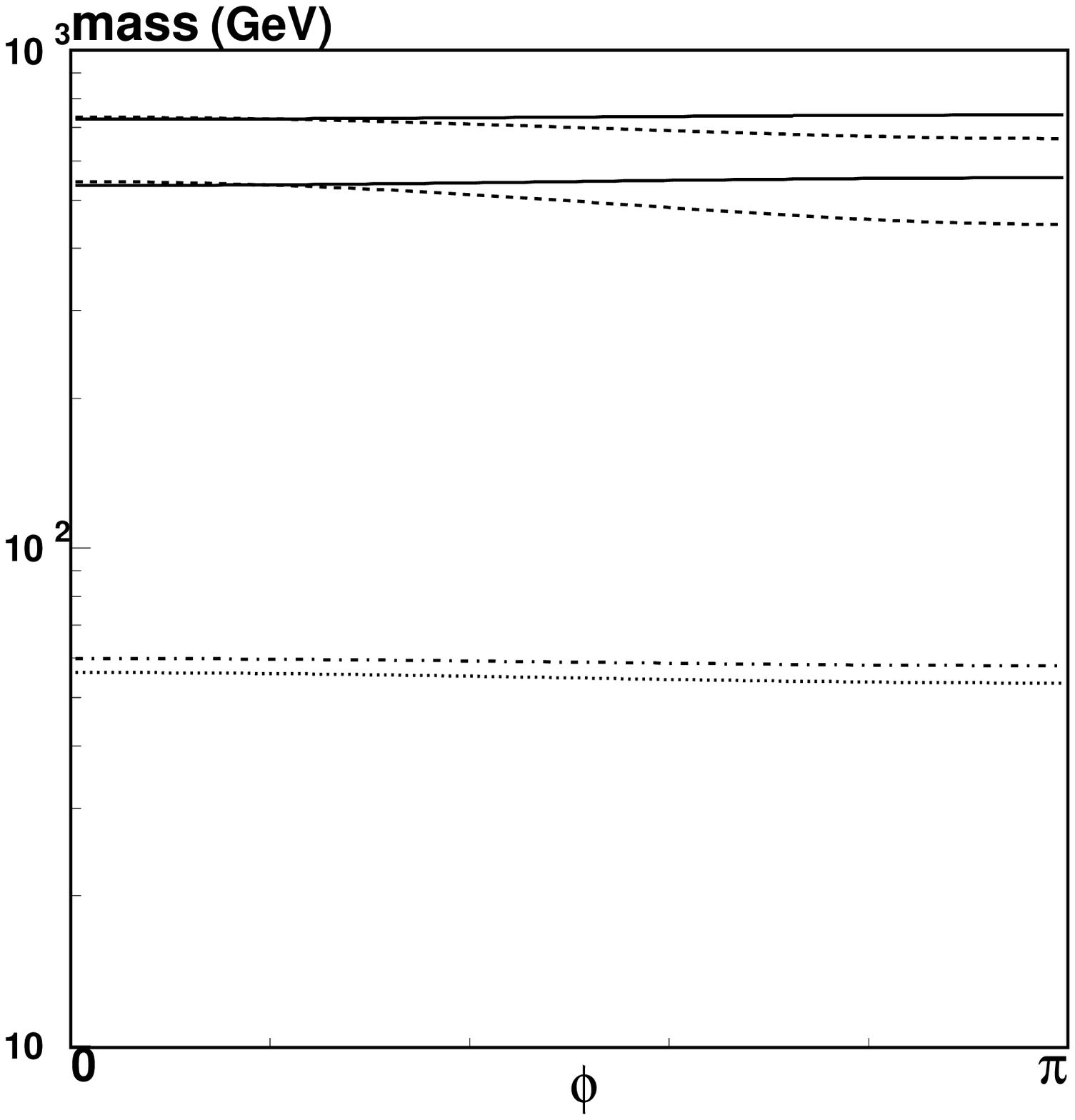}
\caption[plot]{The same as Fig. 2(a), except for $\tan \beta = 10$.}
\end{figure}
% (FIG 3)
\renewcommand\thefigure{FIG. 3.}
\begin{figure}[t]
\epsfxsize=13cm
\hspace*{2.cm}
\epsffile{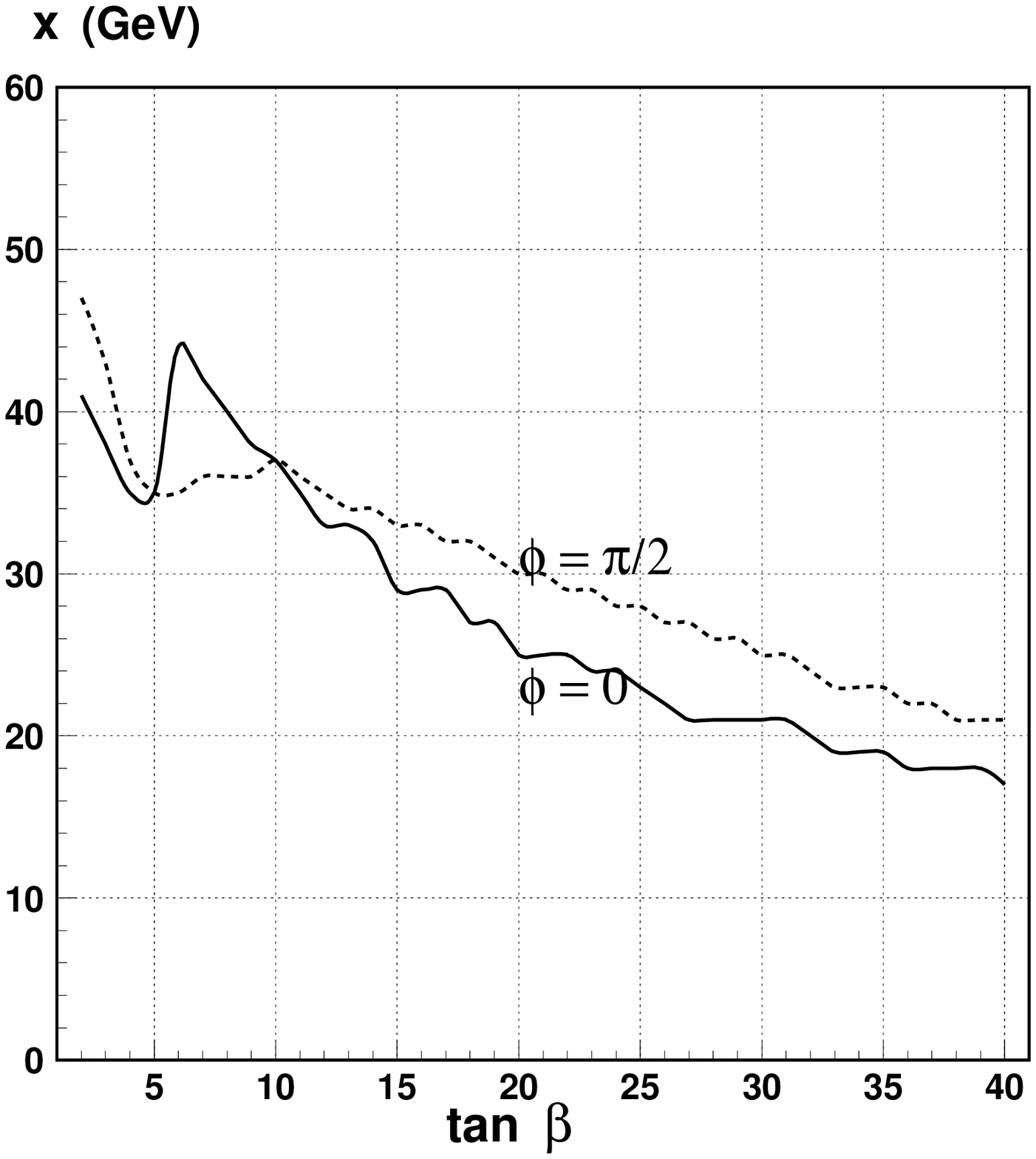}
\caption[plot]{The lower bound on $x$ against $\tan \beta$ under the condition of $\sigma_t < 0.1$ pb for $\phi = 0$ and $\phi = \pi/2$. 
In this calculations, the remaining parameters are independently varied for $0 < \lambda \leqslant 0.87$, $0 < k \leqslant 0.63$, $0 < A_{\lambda}, A_k, m_Q, m_T \leqslant 1000$ GeV, and $0 < A_t \leqslant 2000$ GeV.}
\end{figure}
% (FIG 4a)
\renewcommand\thefigure{FIG. 4. (a)}
\begin{figure}[t]
\epsfxsize=13cm
\hspace*{2.cm}
\epsffile{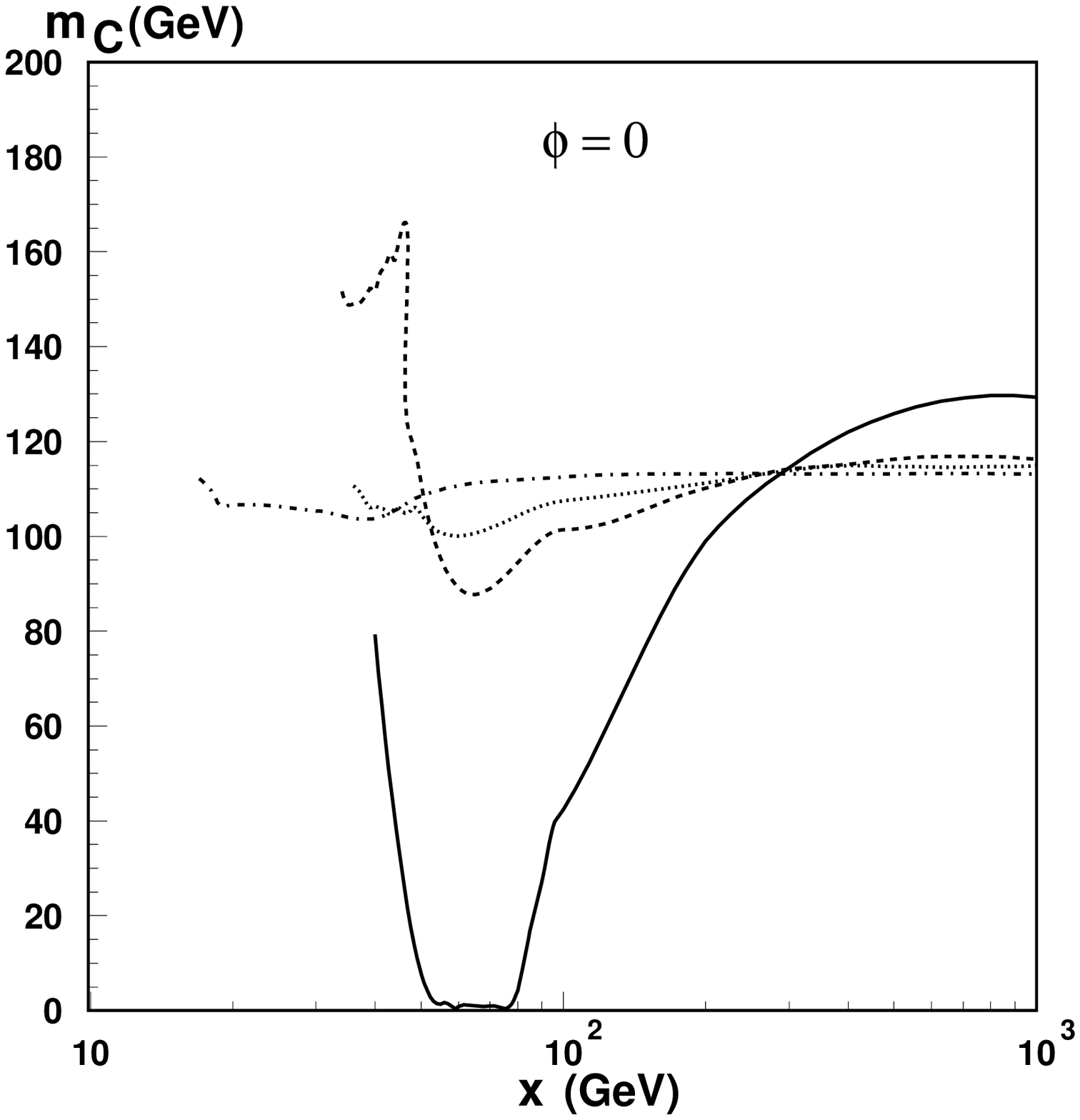}
\caption[plot]{The lower bound on $m_C$ under the condition of $\sigma_t < 0.1$ pb as a function of $x$, for $\tan \beta = 2$ (solid curve), 5 (dashed curve), 10 (dotted curve), and 40 (dot-dashed curve). In this calculation, the remaining parameters are independently varied for the same parameter space as Fig. 3.}
\end{figure}
% (FIG 4b)
\renewcommand\thefigure{FIG. 4. (b)}
\begin{figure}[t]
\epsfxsize=13cm
\hspace*{2.cm}
\epsffile{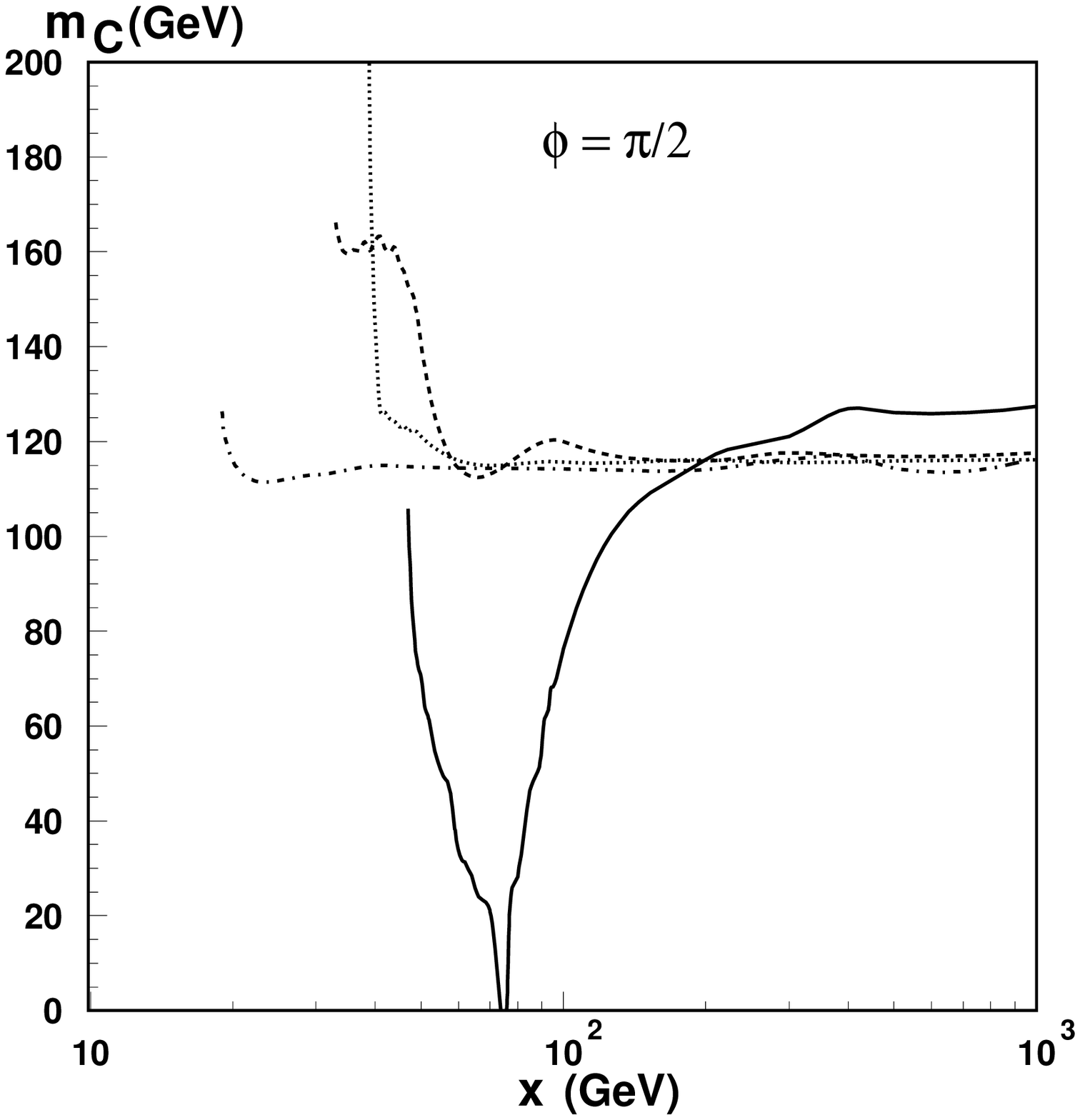}
\caption[plot]{The same as Fig. 4(a) except for $\phi = \pi/2$.}
\end{figure}
%***********************************************************************
\end{document}